\documentclass[a4paper,fleqn,usenatbib]{mnras}

\usepackage[T1]{fontenc}
\usepackage{ae,aecompl}

\usepackage{graphicx}   
\usepackage{amsmath}    
\usepackage{amssymb}    
\usepackage{cases}

\usepackage{amsbsy}
\usepackage{amsfonts}
\usepackage{mathtools}

\usepackage[]{color,hyperref}

\renewcommand{\vec}[1]{\boldsymbol{#1}}
\newcommand{\der}{\mathrm{d}}

\newcommand{\dv}[3][]{\ensuremath{\frac{\mathrm{d}^{#1} #2}{\mathrm{d} #3}}}

\title[NS atmospheres heated by particle bombardment]{Atmosphere of strongly magnetized neutron stars
heated by particle bombardment}

\author[Denis Gonz\'alez-Caniulef et al.]{
Denis Gonz\'alez-Caniulef,$^{1}$\thanks{E-mail: denis.caniulef.14@ucl.ac.uk (DGC)}
Silvia Zane,$^{1}$ Roberto Turolla$^{2}$ and Kinwah Wu$^{1}$\\
$^{1}$Mullard Space Science Laboratory, University College London, Holmbury St. Mary,
Surrey, RH5 6NT, UK.\\
$^{2}$Department of Physics and Astronomy, University of Padova, via Marzolo 8,
I-35131 Padova, Italy
}

\date{Accepted XXX. Received YYY; in original form ZZZ}

\pubyear{2015}

\begin{document}
\label{firstpage}
\pagerange{\pageref{firstpage}--\pageref{lastpage}}
\maketitle

\begin{abstract}

 The magnetosphere of strongly magnetized neutron stars, such as magnetars,
 can sustain large electric currents. The charged particles return to
 the surface with large Lorentz factors, producing a particle bombardment.
 We investigate the transport of radiation  in the atmosphere of strongly magnetized
 neutron stars, in the presence of particle bombardment heating. We solve the
 radiative transfer equations for a  gray atmosphere in the Eddington
 approximation, accounting for the polarization induced by a strong
 magnetic field. The solutions show the formation of a hot external layer
 and a low (uniform) temperature atmospheric interior.   This suggests
 that the emergent spectrum may be described by a single blackbody
 with the likely formation of a optical/infrared excess (below $\sim 1$~eV).
 We also found that the  polarization is strongly   dependent on both
 the luminosity and  penetration length of  the particle bombardment.
 Therefore, the thermal emission   from active sources, such as transient
 magnetars, in which the luminosity   decreases by orders of magnitude, may
 show a substantial variation in the  polarization pattern during the
 outburst decline. Our results may  be  relevant  in view of future X-ray
 polarimetric missions such as IXPE and eXTP.

\end{abstract}



\section{Introduction}
\label{int}

Magnetars are neutron stars (NSs) harbouring  superstrong
magnetic  fields, with strength up to $B\sim 10^{14}-10^{15}$~G
\citep{Duncan92,Thompson95}. Magnetar candidates are usually
associated with   two classes of isolated NSs (now thought to be
unified into a single class), the  Soft Gamma-ray  Repeaters (SGRs)
and the Anomalous X-ray Pulsars (AXPs, for  reviews see
\citealt{Woods06, Mereghetti08, Turolla15,Kaspi17}).
One of the hallmarks of magnetars is the emission of repeated hard
X-ray/soft gamma-ray bursts. Also, magnetars emit a
persistent X-ray luminosity ($L_X \sim 10^{33}-10^{36}~\rm{erg~s}^{-1}$),
that can not be explained in terms of rotational energy loss or
accretion from a binary companion, and is thought to be powered
by their superstrong magnetic field.

Magnetar burst activity has been also observed in few NSs
with  spin-down    (i.e. dipolar)  magnetic  field  of
 $\sim 10^{12}-10^{13}$~G,  which is of the same order
of that found in radio pulsars \citep{Turolla13}. With respect
to standard magnetars, these sources (called ``low field
magnetars'') have longer characteristic age, lower persistent luminosity
and  fewer and less energetic bursts, suggesting that they may
be ``old magnetars'',  in which the external magnetic field
has substantially decayed but the internal magnetic field
is still  strong  enough  to  stress  the  crust  and
hence  produce bursts/outbursts
\citep{Esposito10,Rea10,Turolla11,Rea12}. Indeed,
  X-ray observations of  one low magnetic field magnetar,
SGR 0418$+$5729  which has a  spin-down  magnetic field
$B \approx 6\times10^{12}$~G, showed a phase and energy dependent
absorption spectral feature  consistent with
a  magnetic component with strength of $\sim 10^{14}-10^{15}$~G,
localized near the star surface \citep{Tiengo13}.

Magnetar  persistent  emission  spans  from  the  infrared to
the  soft  gamma-rays. The softX-ray spectra (from $\sim 1$ to $\sim 10$~keV)
typically show a thermal component, presumably  coming
from  the  stellar  surface,  and  a  non-thermal component, probably
produced by resonant cyclotron scattering of thermal photons by energetic
electrons in a twisted magnetosphere (for a review see \citealt{Turolla15}).
These emissions are well modelled by one (or more)  blackbody
component(s) and a power-law component. However, the surface thermal emission
from NSs is expected to be reprocessed by an atmosphere (or, in the case of
the cooler  highly magnetized NSs, by a condensed solid crust;
\citealt{Lai97,Lai01,Burwitz03,Turolla04,Medin07}). Realistic atmospheric
models  are therefore required, in order to model the observed
magnetar spectra in a self-consistent manner.

To solve the transport of radiation in the atmosphere
of a magnetar and compute the emergent spectra, a
number of effects need to be considered.
 First,
the radiation field is  highly polarized: due to strong
magnetic field,   radiation propagates in the atmosphere
in form of two normal modes.  { The so called extraordinary
(X-mode) and ordinary mode (O-mode)  have, respectively,
the wave electric field oscillating in direction perpendicular
and parallel to the plane defined by the wave
vector $\vec{k}$
and the external  magnetic field $\vec{B}$.}
The scattering and free free opacities relative
to the X-mode  are,  in  general,  substantially
lower  than  the  ones relative to the ordinary mode.
Indeed, below the electron cyclotron
frequency $\omega_{e,c}$,   the X-mode opacity
is reduced by a factor of order $\omega^2/\omega_{e,c}^2$
with respect the O-mode opacity \citep[see e.g.][]{Harding06}.
 The main consequence is that  the emergent
 spectra are expected to show a large degree of polarization.
 However, quantum  electrodynamic effects  such as vacuum
 polarization, i.e., the creation of virtual $e^{+}e^{-}$
 pairs induced by the strong magnetic field, can introduce
 additional changes to the radiative opacities. In particular,
 when vacuum and plasma effects cancel each other (i.e. at the
 location of the so--called ``vacuum resonances''), resonant
 features in  the opacities and  adiabatic photon conversion
 from one polarization mode to another may occur,
 which in turn modify the transport of radiation
 \citep{Gnedin78,Pavlov79a,Meszaros79,Lai02}.

Several works have addressed the problem of modeling
the transport of radiation in magnetized NS atmospheres.
Self-consistent models for fully ionized atmospheres
with  magnetic fields  $B\sim 10^{12}-10^{13}$~G were
presented by \cite{Shibanov92}  and  \cite{Pavlov94}.
Models for NSs atmosphere with similar magnetic
fields but considering
energy deposited by slowly-accreting matter
have been studied by \cite{Zane00}. Further
models for passively cooling NSs with  stronger fields,
$B \geq 10^{14}$~G were computed by  \cite{Ho01} and
\cite{Ozel01}.  The problem of  cyclotron line  creation
in the emergent spectrum  was first  addressed  by
\cite{Zane01} and  then by   \cite{vanAdelsberg06},
 who showed a potential suppression of this feature due
 to mode conversion at the vacuum resonance.  Partially
ionized atmospheres with realistic equation of state
were investigated by \cite{Ho03b} and \cite{Potekhin04a},
and later, mid-Z elements atmospheres with
$B\sim10^{12}-10^{13}$~G by  \cite{Mori07}.
Thin hydrogen atmospheres above either a condensed surface
or helium layer were studied by \cite{Ho07} and
\cite{Suleimanov09}, respectively.
 The effects of cyclotron harmonics were investigated
 by \cite{Suleimanov12}.
 A recent detailed review can be found in \cite{Potekhin14}.

What limits the application of the existing cooling
models to the case of magnetars is that they are under
the assumption that the NS is a ``passive cooler'', i.e. that
the only heat source in the atmosphere originates from energies
from the  stellar  core.
This is quite an unrealistic simplifications for sources
like magnetars, which are characterized by strong
magnetospheric activity and are believed to possess
complex magnetospheric configurations.
For instance, in magnetars the external magnetic field
is expected to be twisted. In fact,   according to the
magnetar model \citep{Thompson02}, the star internal magnetic
field has both a  toroidal  and a  poloidal  component.
 As  the  magnetic  field  evolves, strong Lorentz forces and
 plastic deformations will develop in the stellar crust.
 They can transfer helicity from the internal toroidal field to
 the external magnetic field, producing a twisted magnetosphere.
This non-potential, twisted magnetic field is
sustained by large currents  of charged particles, that in turn move
along the twisted field lines and can  return toward the star
surface, depositing energy in the atmosphere, raising the gas
temperature and affecting the transport of radiation (``particle
bombardment'' effect).

Twisted  magnetospheres  are  not  equilibrium
configurations,  and  they  are  expected  to  relax.
Indeed,  after  the sudden injection of helicity
following the crustal motion, the magnetosphere should
gradually untwist.
 According  to \citet{Beloborodov09},
two  regions coexist in a untwisting  magnetosphere: (i) a
cavity, that  is described by a potential  region where
$\nabla \times B =0$, and (ii) a twisted region, where
 $\nabla \times B \neq 0$.  As the magnetosphere untwists,
 the cavity expands and the footprints of the twisted field
 create a hot spot at the magnetic pole that shrinks over time.
 This means that the portion of the atmosphere affected by
energy deposition by the inflowing magnetospheric particles,
and the amount of heat associated with particle bombardment,
is also expected to change as the magnetosphere untwists.

In this work, we present a model for the thermal emission
from magnetized NSs under particle bombardment. We study the transport
of radiation, in an atmosphere heated by a back-flow of charged particles
from the magnetosphere. 
{ The deceleration of these particles can produce an $e^+e^-$ pair cascade 
that  can deposit energy in the  atmosphere interior.
We organise the paper as follows.
In \S\ref{stopping} we discuss the deceleration
process/cascade formation,  and calculate the associated stopping length.}
In \S\ref{radiative}, we present the model radiative
transport equations under the conditions of hydrostatic
equilibrium and energy balance. We consider the case of a ``grey''
plane-parallel atmosphere. We solve the transport equations
in the Eddington  approximation  and  obtain  the  intensities
for  the  two polarization modes. We present our results and implications
in \S\ref{res} and conclusions  in
\S\ref{conclusions}.

\section{Theoretical framework}
\label{theo}

\subsection{Stopping length}
\label{stopping}

 Fast, returning,  magnetospheric particles decelerate
and deposit energy as they penetrate the atmosphere.
In the case of an unmagnetized atmosphere, the stopping
length for fast electrons is dominated by relativistic
bremsstrahlung (\citealt{Bethe34,Heitler54,Tsai74}, see also
\citealt{Ho07}).  For 
moderate  magnetic fields,  below  $B\sim10^{13}$~G, fast 
particles lose their kinetic  energy mainly due to 
Magneto-Coulomb  interactions\footnote{Collisionless processes 
such as beam-instability (e.g., \citealt{Godfrey75}) have 
been also proposed to  contribute to the stopping length, although  
detailed calculations still need to be done 
(see \citealt{Thompson05,Beloborodov07}). However, \citet{Lyubarsky07}, 
have shown that collisionless dissipation does not work in
the atmosphere of a NS because the two-stream instability 
is stabilized by the inhomogeneity (density gradient) of the atmosphere.},
and the corresponding cross sections have 
been derived by \citet{Kotov85,Kotov86}.
Since the  particle stopping length in the case of  
strongly  magnetized  plasma, i.e. 
$B \gg 10^{13}$~G, is still  not very  well understood,
the cross sections computed by \citet{Kotov85,Kotov86}
 have also been used 
by \citet{Lyubarsky07} in order to derive approximated results for the 
deceleration of fast particles at larger field strengths ($B\approx 10^{14}-10^{15}$~G).

The process can be explained as follows. After impinging onto the
atmospheric layer,  returning  magnetospheric  particles
move along the magnetic field  lines, while occupying the fundamental
Landau level. After a scattering with a plasma nucleus, the
fast particle can make a  transition to an excited Landau level.
If  the  timescale for  collisions between fast particles
and  plasma nuclei is  larger than the  de-excitation timescale to return
to  the  fundamental  Landau level, then  a high  energy photon is emitted  and
the fast particle returns to  the  fundamental  Landau level
before experiencing a new scattering. In the following we will
refer to this regime as ``low density plasma'', which occurs when
\citep{Kotov85}
\begin{equation}
 c\sigma_{01}n \ll \frac{1}{\gamma \tau},
\label{eq:mc_condition}
\end{equation}
where $c$ is the speed of light, $\sigma_{01}$ is the
cross section for exciting the charged particle to the first
Landau level, $n$ is the number density of nuclei, $\gamma$
is the Lorentz factor of the impinging particle, and $\tau$ is the
life--time of the first excited Landau level.

\begin{figure*}
\centering
\includegraphics[width=16.0cm,viewport=0 0 900 850,clip]{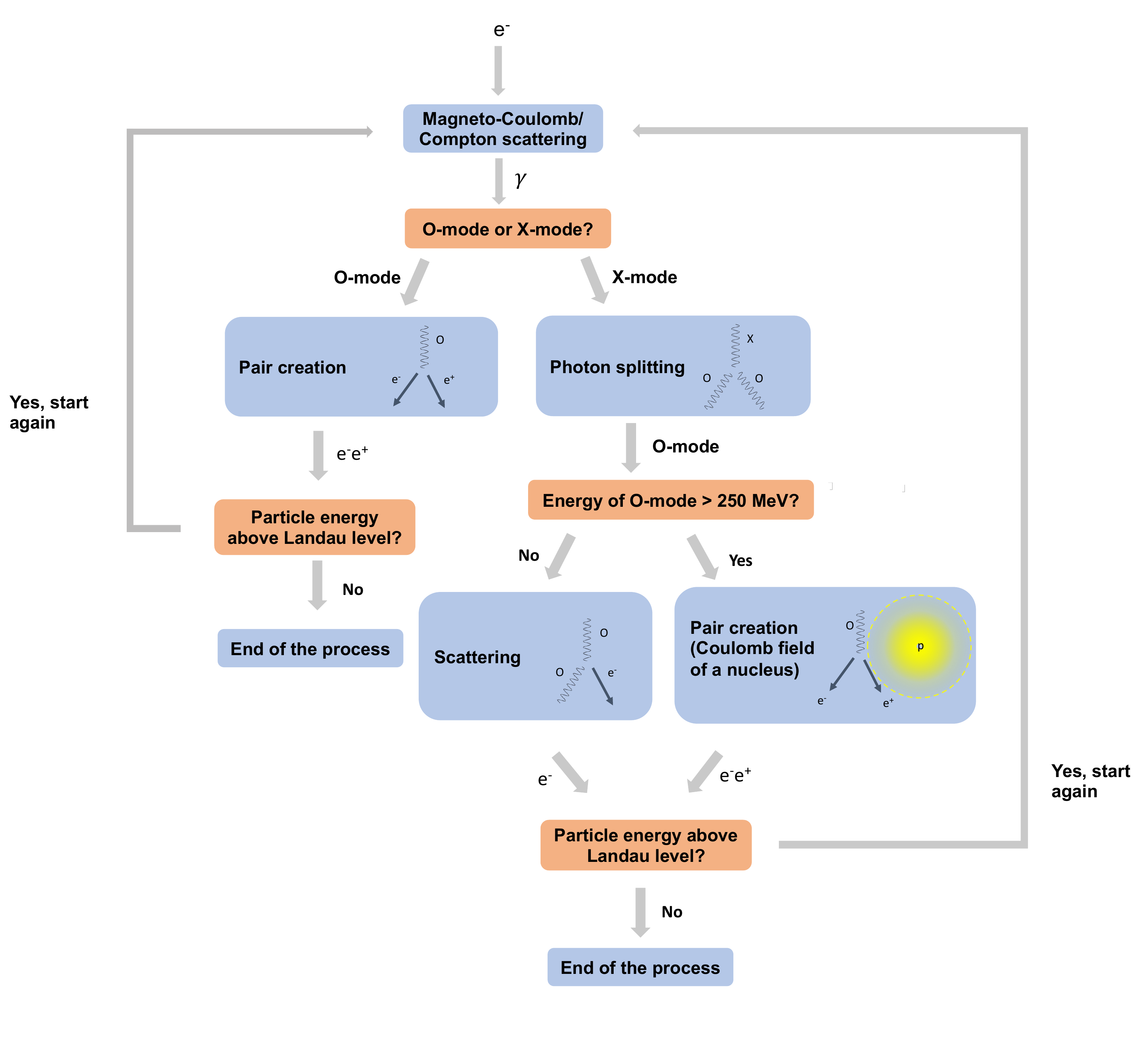}
\caption{{Electron-positron  avalanche due to particle bombardment.
 The figure refers to magnetic field  $B\sim 10^{15}$~G, in which
case, after the initial Magneto-Coulomb or Compton scattering interaction  most of the energy is transferred to 
photons (either O- or X-mode) and the electron retains a limited fraction of its kinetic energy.
The long, right hand side branch that departs after the Magneto-Coulomb/Compton Scattering interaction 
describes  the different mechanisms triggered by the initial penetration of a  X-mode photon (under
the assumption that O-mode photons resulting from the photon splitting have energy 
below the  threshold for direct pair creation). In this branch, after a scattering,  the O-mode photon transfer most
of its energy to an electron. See text for details}}
\label{fig:avalanche}
\end{figure*}

As discussed by \citet{Lyubarsky07}, during collisional Landau level
excitations most of the energy is transferred to photons (the seed
electrons retain only a small fraction of the total energy). The
resulting gamma-rays can trigger an electron-positron pairs avalanche,
ultimately leading to electrons with the energy below the Landau energy.
If the pair avalanche is triggered by O-mode gamma rays, the typical
length scale in which it occurs is roughly coincident with the
stopping length of the charged particles and is given by
\citep{Lyubarsky07}
\begin{equation}
 l_{_{Low}} \approx \frac{2B}{3 \xi B_q Z^2 n_i \sigma_T}\ln
 \left [ \ln \left(\frac{0.4 \gamma^2 B_q}{B} \right) \right ] \, ,
\label{llow}
\end{equation}
where $\xi \approx 0.15 (B/10^{15}~\rm{G})^{-1/2}$ is
the fraction of the kinetic energy retained by the electron
after one scattering, $B_q=m_e c^3/\hbar e = 4.4\times 10^{13}$~G
is the critical quantum field, $Z$ is the atomic number,
$\sigma_T$  is the Thomson cross section, and $n_i$
is the number density of ions.
For $\gamma \sim 100-1000$, this translates into a Thomson depth of
\begin{equation}
\tau_0 = \sigma_T Z n l_{Low} \approx 200 Z^{-1} \left ( \frac {B}{10^{15} \, {\rm G}} \right )^{3/2} \, .
\label{eq:taulow}
\end{equation}

\begin{figure}
\centering
\includegraphics[width=8.0cm,viewport=20 5 550 400,clip]{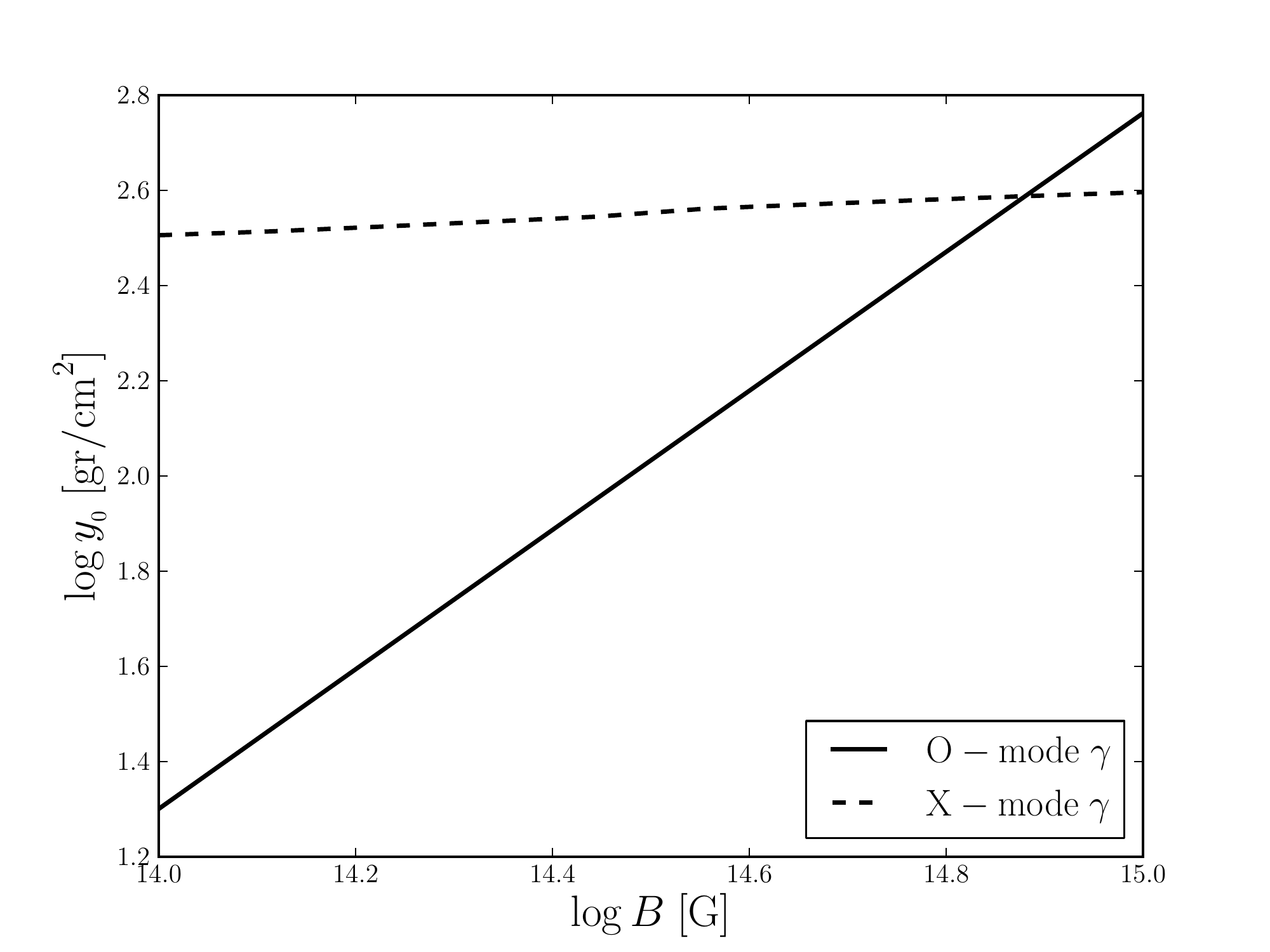}
\caption{Stopping column density for different magnetic fields.
The solid line shows  the penetration length of the
electron-positron avalanche produced by the initial
O-mode photon branch  (after a magneto-Coulomb interaction
for  particles with Lorentz factor $\gamma = 10^3$).
The dashed line shows the penetration of the electron-positron
avalanche due to X-mode photons.}
\label{fig:stopping}
\end{figure}

X-mode gamma rays do not produce pairs directly  but 
they first split into O-mode photons (see the right hand side branch
after the initial Magneto-Coulomb interaction in  Figure \ref{fig:avalanche}).
If the energy
of the resulting O-mode photon is below the energy threshold
for direct pair creation then {\bf a)} pairs may be created via
the interaction with the Coulomb field of a nucleus or 
{\bf b)} the O-mode photon may transfer the energy to an
electron via Compton scattering, which then may undergo a 
Magneto-Coulomb interaction.
\cite{Lyubarsky07} calculated the mean free path  for successive
production of X-mode photons for particles with Lorentz factor $\gamma=10^3$,
moving in a magnetic field of $B=10^{15}$~G, finding that 
the avalanche can extend up to a relatively large optical depth 
\begin{equation}
\tau_0  \approx 300 Z^{-1}.
\label{eq:taulowtotal}
\end{equation}
In a  hydrogen atmosphere, this translates into a relatively large
column density for stopping the successive production of  X-mode photons
$y_{_0} \sim 750~\rm{g~cm}^{-2}$,  while seed electrons from the particle 
bombardment are stopped at  $y_{_0} \sim 570~\rm{g~cm}^{-2}$.
 However, by repeating
the same calculations for a magnetic field $B=10^{14}$~G, we found
that the stopping column density  can be reduce to $y_{0} \sim 250 ~\rm{gr~cm}^{-2}$
for X-ray photons, and to $y_{0} \sim 20 ~\rm{gr~cm}^{-2}$ for seed
electrons\footnote{Notice that in a magnetized vacuum with
$B\sim 10^{14}$~G, pair creation triggered by  X--mode photons can 
also occur (see threshold in section 5 of \citealt{Weise2006}). However,
this process may change in presence of a magnetized plasma
and no computations for the corresponding cross sections are available.
In the following, we do not take this effect  into further account.}.
It should be noticed that,  in the case in which the energy of the  O-mode photon is above the
threshold for direct pair creation,   the pair avalanche penetrates the 
atmosphere up to a depth  significantly  less than that of equation 
(\ref{eq:taulow}).
The stopping column densities for different magnetic fields for both the X-mode
and O-mode branches are also shown in  Figure \ref{fig:stopping}.

In a ``high'' density plasma, i.e. such that
 $c\sigma_{01}n \gg 1/(\gamma \tau)$,
the impinging charged particles can be  excited,
via successive scatterings, to Landau levels with large
quantum numbers \citep{Kotov86}. In this case, the electron
mean free path has been calculated by
\citet{Kotov86}, and is given by
\begin{equation}
 l_{_{High}} = 2.35\times10^8 \frac{B_q}{ZB}
 \left(\frac{\xi}{n \Lambda}\right)^{1/2},
 \label{eq:stop}
\end{equation}
where $\Lambda$ is the Coulomb logarithm.

 \citet{Potekhin14} compared this result with the geometrical depth obtained from
the NS envelope models of \citet{Ventura01} and performed 
a numerical fit to find a  simple  analytic formula
for the column density at which particles decelerate. 
Assuming
that $\Lambda=1$, this column density is given
by\footnote{Note that this expression differs  from that presented
by \citet{Potekhin14}, since it contains corrections discussed in a
private communication with Dr. Potekhin.}
\begin{equation}
 y_{_0,High} \approx \left[ \left(\frac{A}{Z^2} \frac{\gamma}{700}\right)
 \left(\frac{B}{10^{12}~\rm{G}}\right)^{-2}\right]^{0.43}
\left(\frac{T}{10^6\rm{K}}\right)~\rm{g~cm}^{-2},
\label{eq:pot14}
\end{equation}
where $T$ is the temperature of the atmosphere, which enters in
the expression of the stopping length because a definite atmospheric
model was used.
It should be noticed that Eq. (\ref{eq:pot14}) scales with the magnetic
field as $\propto B^{-0.86}$, which
means that for stronger  magnetic fields the column density
decreases, as well as the density at which particles
lose most of their kinetic energy. Then, for strong  magnetic
fields, large Lorentz factors are required in Eq. (\ref{eq:pot14})
to satisfy the criterion of high density plasma.  It  is worth noticing also
that, since the low stopping length inferred by Eq. (\ref{eq:pot14})
have been invoked to suggest the existence of a self-regulating mechanism
driven by magnetospheric currents which may
produce thin hydrogen layers on the surface of ultra magnetized NSs
\citep{Ho07,Potekhin14}, the existence of such ultra-relativistic particles
($\gamma \sim 10^{14}$ for $B\sim 10^{13}$~G) is necessary in order to
make the scenario self--consistent.

In a twisted magnetosphere, particles are accelerated to
typical { Lorentz factor $\gamma \sim 10^3-10^6$. Assuming
$\gamma=10^3$, and considering} magnetic fields $B\sim 10^{14}$~G, the
cross section to excite
particles to the first Landau level is
$\sigma_{01} \sim 10^{-25}~\rm{cm}^2$ (assuming $\Lambda=1$), and the
typical life-time of the process is $\tau \sim 10^{-19}~\rm{s}$ \citep{Kotov85}.
This means that  the condition of low density plasma (Eq.\ref{eq:mc_condition}) requires
 $n \ll 10^{29}~\rm{cm}^{-3}$,
which is  typically met across the NS atmosphere. 
Thus, in the following, only the stopping length corresponding 
to the low density case is considered.
 We should notice that the
estimates for the stopping optical depth, eqs. (\ref{eq:taulow})
and (\ref{eq:taulowtotal}) are based on cross sections for
magneto-Coulomb interactions, photon splitting induced by the the
magnetic field (or the Coulomb field of a nucleus), and Compton
scattering processes (Klein-Nishina cross section) that are not
accounting for all the complex physics of plasma in  super
strong magnetic fields. Detailed  calculations for these processes
in the magnetar regime are outside the scope of this study. {
However, we can conclude that by considering a range of stopping
column densities as large as $y_{_0} \approx  50
-500~\rm{g~cm}^{-2}$, gives scenarios fairly compatible with
values previously discussed for particles decelerated in a low
density plasma, inasmuch the lower limit is representative  of the
penetration of the  particle bombardment for magnetic fields
$B\sim 10^{14}$~G, while the upper limit is representative of the
penetration of successive X-mode photons production for $B\sim
10^{15}$~G. Both cases assume initial particle bombardment with
Lorentz factor $\gamma = 10^3$.

Another important process that can contribute to the stopping
length is resonant Compton drag. In fact, upon impinging on the
atmosphere, electrons feel a hot photon bath (with typical
temperature of $ T \sim 10^7$~K, see next sections) and, in presence
of a high magnetic field, may undergo significant resonant Compton
cooling. This process has been investigated by \cite{Daugherty89},
 \cite{Sturner95} and, more recently, by \cite{Baring11} who calculated
the electron cooling rate using the fully relativistic, quantum
magnetic Compton cross section. As shown by \citet[][see their
Fig.~6]{Baring11}, the electron cooling length scale varies
considerably with both, the Lorentz factor of the charges and the
photon temperature and reaches a minimum at the onset of the
resonant contribution, in correspondence of $\gamma_r \approx
B/(\Theta B_q )$, where $\Theta = kT/m_ec^2$. They found that, for
$B\gg B_q$, the absolute minimum of the Compton stopping length is
\begin{equation}
\lambda_C \sim \frac{2 \tilde \lambda B}{B_q \alpha_F \Theta^3}\,,
\end{equation}
where $\tilde \lambda$ is the reduced electron Compton wavelength,
and $\alpha_F$  the fine structure constant. The previous
expression strongly depends on the temperature.

 For $B=10^{14}$~G and $T=10^7$~K, this gives $\lambda_C \sim 100~\mathrm{cm}$, in
correspondence of a Lorentz factor $\gamma \sim 10^{3}$. In turns, assuming
a density $\rho\sim 10^{-3}~\mathrm{g~cm}^{-3}$, this translates
into a stopping column density as small as 
$y_0 \sim 0.1~\mathrm{g~cm}^{-2}$. Indeed, by assuming that the Compton stopping
length grows linearly with $\gamma$ for $\gamma > \gamma_r$  
\citep[see the curves in Fig.~6 of][]{Baring11}, we can estimate that
the Compton stopping length becomes larger than the stopping
length for magneto-Coulomb collisions (so the latter effect
becomes dominant) only if the charge Lorentz factor is $\gamma> 10^6$.

On the other hand, even considering $\gamma\sim 10^3$, we stress that
an expected value $y_0 \sim 0.1~\mathrm{g~cm}^{-2}$ is an absolute
lower limit, for several reasons. First, it assumes monoenergetic
particles at $\gamma_r$. As soon as the Lorentz factor departs
from this value, the resonant process becomes more inefficient and
the Compton stopping length increases (extremely rapidly at lower
$\gamma$, and almost linearly for $\gamma > \gamma_r$). Moreover,
this value is estimated assuming that the radiation field is
isotropic (while this is not expected to be the case in the
external layers of our magnetized atmospheres) and the resonant
conditions (which is in general angle-dependent) is met by all
impinging charges.

In any case, while experiencing Compton drag, the cooling
electrons produce hard X-ray/soft gamma-ray photons, primarily but not
exclusively in the X mode \citep{Beloborodov13,Wadiasingh18}, and
these X-mode photons can split into O-mode photons, which  can produce
a pair cascade, contributing to
the process previously described (see discussion after Eq.~3). In
order to estimate the stopping length at which the pair cascade
(and heat deposition) end, we can then  repeat the same
calculation as before, by using again a magnetic field $B =
10^{14}$~G. If we assume that a seed electron (stopped at
 $y_0=0.1~\mathrm{g~cm}^{-2}$) with Lorentz factor $\gamma = 10^3$ transfers all the
energy, or 1/2 of the energy, or 1/4 of the energy  to a
single X-mode photon, then the results is similar to what
discussed before, i.e. the  e$^+$e$^-$ cascade stops at 
 $y_0 = 326~\mathrm{g~cm}^{-2}$, $y_0 = 321~\mathrm{g~cm}^{-2}$
    or $y_0 = 187~\mathrm{g~cm}^{-2}$, respectively. This is because, during the
cascade, the stopping column density is dominated by the characteristic
length travelled by O-mode photons (produced by splitting of X-mode photons) 
before producing pairs, rather than the path travelled by the pairs themselves  
(which can be effectively  stopped by resonant Compton drag at the site of creation).

In the following, we will carry on the analysis by considering a
range of stopping column densities 
$y_{_0} \approx  65-500~\rm{g~cm}^{-2}$, which covers the scenarios described before
for charges with impinging Lorentz factor $\sim 10^3$. In this
range of values, we ensure a convergence of our numerical code.
The potential effects of a larger Lorentz factor are briefly
discussed in the conclusions and summary section.}

For simplicity, we assume that  the heat deposition
is distributed uniformly along this depth, and is given by
\begin{equation}
W_{_H} = L_\infty/4 \pi R^2 y_{_0}
\label{eq:heat}
\end{equation}
where $L_\infty$ is the luminosity at infinity and $R$ is the star
radius. This expression is valid as long as
 the velocity, $v$, of incoming magnetospheric particles is
much higher than the thermal velocity of the plasma, $v_{th}$, which
allows us to ignore  collective  plasma oscillation \citep{Alme73,Bildsten92}.
We assume that the only heating source
in the atmosphere is due to the particle bombardment, which means we
set $W_H=0$ for $y> y_{_0}$.

\subsection{Radiative transfer}
\label{radiative}

In the following  we compute the transport of radiation  in magnetized atmospheres
heated by  particles bombardment induced by a  twisted
magnetosphere. 
According to \cite{Beloborodov07}, the twist current 
$\vec{j} = (c/4\pi) \nabla\times\vec{B}$ is maintained by a voltage
$\Phi$ that is regulated by a $e^+e^-$ discharge along the magnetic
field lines.
For  a non-rotating NS with a twisted dipole, as the 
magnetosphere untwist, the cavity in which $\nabla\times\vec{B}=0$
expands out from the equatorial plane,  and the currents become concentrated
in a  ``j-bundle'' which footprints are anchored near the magnetic poles.
The evolution  of the j-bundle, where the back bombarding 
currents are accelerated, produces a  polar hot spot that shrinks over time.
The luminosity associated to currents returning onto the NS surface is
given by \citep{Beloborodov09}.
\begin{equation}
 L \approx 1.3 \times 10^{36} B_{14} R_6 \psi {\cal V}_9  u_\star^2 ~
\rm{erg~s}^{-1},
\end{equation}
 where $\psi$ is twist angle,  ${\cal V}_9$ is a threshold voltage in units of
$10^9$~V, and 
\begin{equation}
u_\star (t) \approx \frac{cR{\cal V}}{2\mu\psi} (t_{end} -t)
\end{equation}
 defines the (time dependent) magnetic flux surface in which is located
the current front (or sharp boundary of the cavity), with 
$t_{end} = \mu \psi_{_0}/cR{\cal V}$ the time required to erase the twist.
 Here,  $\psi_{_0}$ is the amplitude if the twist imparted by the starquake 
 and $\mu$ is the magnetic dipole moment. Owing to the gravitational redshift, the
total luminosity seen by a distant observer, $L_\infty$, is
related to the local luminosity at the top of
the atmosphere by $L_\infty = y_{_G}^2 L(0)$,  where
$y_{_G}=(1-R/R_s)^{1/2}$ is the gravitational redshift factor in a
Schwarzschild spacetime for a star
with radius $R$ and Schwarzschild radius $R_s$.

The envelope material is assumed to be pure, completely ionized
hydrogen.
In order to solve the energy balance and radiative transfer,
we work in plane-parallel approximation and we compute equilibrium
solutions in the frequency-integrated case. { We follow the
same approach that has been used by \cite{Turolla94},
\cite{Zampieri95}, and  \cite{Zane00}
for computing  model atmospheres at low accretion rates in the
non-magnetized and moderately magnetized regime \citep[see also][for earlier works]{Zeldovich69,Alme73}.}
The structure of the atmosphere is obtained by solving
the hydrostatic equilibrium equation
\begin{equation}
\dv{P}{y} = \frac{GM}{ y_{_G}^2 R^2 },
\label{eq:hydrostatic}
\end{equation}
 where   $G$ is the  gravitational constant, $M$ is the mass of the star
and $P = k \rho T/\mu_e m_p$ is the gas pressure, with $T$ the gas temperature and
$\mu_e=1/2$ (for fully ionized hydrogen). 
Eq. (\ref{eq:hydrostatic}) can be solved analytically, giving
the density as a function of the column density
\begin{equation}
\rho = \frac{GMm_p}{2y_{_G}^2R^2}\frac{y}{k T(y)} \, .
\label{eq:density}
\end{equation}
We are interested in luminosities typical of magnetar sources,
$L = 10^{34}-10^{36}~\rm{erg~s}^{-1}$, in which case both,
the ram pressure by the particle bombardment  and the
radiative force  are negligible with respect to the thermal pressure and the
gravitational force, respectively.

The differential equation for the total luminosity is given by
\citep{Zampieri95}
\begin{equation}
\frac{1}{4\pi R^2}\dv{L}{y} =  - \frac{W_H}{y_{_G}},
\end{equation}
which can be combined with Eq. (\ref{eq:heat}) to derive an analytical solution
for the luminosity,
\begin{numcases}{L(y)=}
 \frac{L_{\infty}}{y_{_G}} \frac{y_{_0}-y}{y_{_0}} & $y<y_{0}$\label{eq:luminosity} \\
   0 & $y \geqslant y_{_0}$ \nonumber
\end{numcases}

In a  magnetized plasma, the polarized radiative transfer problem
can be described by a  system of two coupled differential
equations for the specific intensities of ordinary and extraordinary
photons. In the Eddington and plane-parallel approximation,  the first  and second
moments  of the radiative transfer equations can be written
in terms of the  luminosity, $L_i$,  and radiation energy
density, $U_i$, of each normal  mode (with $i=1,2$), both
measured by the local observer, as
\begin{eqnarray}
\label{eq:transfer1}
\frac{1}{4\pi c R^2}\dv{L_i}{y} &=&    \kappa^P_i
\left(\frac{a T^4}{2} - U_i\right) + \bar S (U_{3-i} - U_i),\\
\frac{1}{3}\dv{U_i}{y} &=&   \kappa^{_R}_i \frac{L_i}{4\pi R^2 c y_{_G}},
\label{eq:transfer2}
\end{eqnarray}
where $\bar S$ is the scattering mean
opacity from one normal mode to the other, and $\kappa_i^P$ and
$\kappa_i^{_R}$ correspond to the Planck and Rosseland mean opacity for
each normal mode, respectively. In absence of a complete
frequency-dependent calculation (which is outside the purpose of this
paper), we approximated the absorption mean opacity and the flux mean
opacity with $\kappa_i^P$a and $\kappa_i^{_R}$, respectively, for each
mode of propagation.

In the following, we neglect collective plasma effects and
consider only the limit $\omega_p^2/\omega^2 \ll1$,
where $\omega_p$ is the plasma frequency. We also consider
only frequencies lower than the electron cyclotron frequency,
so the semitransverse
approximation can be assumed to hold. With the exclusion of
the very external layers, the temperature in the atmosphere is
always lower than $\sim 10^7$~K, so that scattering dominates
over true absorption only for $\tau \sim 1$  and Comptonization
is negligible. For this reason, similarly to what has been
done in the previous investigations, only conservative scattering is accounted for in the
transfer equations.  We consider a plasma composed of protons, electrons
and we do take into account for vacuum corrections in the polarization
eigenmodes, that start to be important when the magnetic field approaches
the quantum critical limit, $\sim B_q$. These corrections induce the
breakdown of the normal mode approximation at the mode collapse points
(MCPs), where the two normal modes can cross each other or have a close
approach, depending on the photon propagation angle (see Pavlov \&
Shibanov 1979, \citealt{Zane00}. see \S\ref{opa}). The detailed
expressions of the magnetized free free and Thomson scattering opacities
are discussed later on, together with the method we
followed for their  numerical evaluation and average (\S~\ref{opa}).

Despite Comptonization is expected to be negligible as
far as the radiative transfer equations are concerned,
in analogy to the case of accreting atmospheres,
we expect that it has an important role in establishing
the correct energy balance in the external, optically thin
layers \citep{Alme73,Turolla94,Zampieri95,Zane00} where the
heating deposited by the incoming charges needs to be balanced
by Compton cooling.  Accordingly, we write the
energy balance equation as
\begin{equation}
\frac{\kappa^P}{\kappa_s} \left( \frac{aT^4}{2}
- \frac{\kappa^P_1}{\kappa^P} U_1 -
\frac{\kappa_2^P}{\kappa^P}  U_2 \right) +
(\Gamma - \Lambda)_c  = \frac{W_H}{c\kappa_s}
\label{eq:balance}
\end{equation}
where $(\Gamma - \Lambda)_c$ is the Compton
heating-cooling term, evaluated using the approximation of  \citet{Arons87},  $\kappa^s$ is
the total (summed over the two modes) scattering opacity and
$\kappa^P = \kappa^P_1 + \kappa^P_2$ is the
total Planck mean opacity.
 One complication is that the Compton heating-cooling term in equation (\ref{eq:balance}) 
depends on  the radiation temperature, $T_\gamma$, which in turn
should be obtained by solving the frequency-dependent  
transport of radiation. To overcome this
problem, we assume that the energy exchange between photons
and the gas particles, due to multiple Compton scatterings,
is governed  by the equation \citep{Wandel84,Park89,Park90,Turolla94,Zampieri95}
\begin{equation}
\frac{y}{T_\gamma} \dv{T_\gamma}{y} = 2 Y_c \left(
\frac{T_\gamma}{T} -1 \right),
\label{eq:compton}
\end{equation}
where
$Y_c~=~(4kT/m_e c^2)\rm{max}(\tau_{es},\tau_{es}^2)$ is
the Comptonization parameter and $\tau_{es}~=~\kappa_{s} y$
is the scattering  optical depth.

 The system of differential equations given by equation \ref{eq:transfer1} 
(solved for mode 1  only, since the luminosity of mode 2 
can be then obtained by difference using equation \ref{eq:luminosity} for the
total luminosity), 
equation \ref{eq:transfer2} (solved for both modes) and equation \ref{eq:compton} 
needs to be solved coupled to a  set of 4 boundary conditions.
First,  we assume a non-illuminated atmosphere, which translates
into a free streaming condition relating the total energy density of the
radiation to the total luminosity, at the outer layer  of the atmosphere,
\begin{equation}
\nonumber
 U(y_{min}) = U_1(y_{min}) + U_2(y_{min}) = \frac{L^\infty_1+
L^\infty_2}{2\pi R^2 c y_{_G}},
\end{equation}
where $y_{min}$ correspond to the minimum value of the column
density grid used
in the atmosphere model.
Second, we assume energy equipartition  in the
inner, optically thick,  atmospheric region, where the energy
density of each mode approaches $\sim aT^4/2$, and we set
$U_1 (y_{max}) = U_2 (y_{max})$, where $y_{max}$ is the maximum
value of the column density grid used in
the model calculation.
Furthermore, we assume photons
in thermodynamic equilibrium with the gas in the inner layers
of the atmosphere, therefore the radiation temperature must be 
equal to the gas temperature  $T_\gamma(y_{max}) = T (y_{max})$.
Finally, since the observed luminosity is  produced by re-irradiation of the 
heat deposited by particle bombardment, with $W_H=0$ 
for  $y > y_{_0}$,
the condition  $L_1(y > y_{_0}) = L_2(y>y_{_0}) =  0$ is used in the 
inner regions of the atmosphere.

\subsection{Opacities}
\label{opa}

Absorption and scattering opacities in a magnetized medium
were widely investigated in the literature (see e.g. \citealt{Meszaros92},
\citealt{Harding06}, \citealt{Potekhin14} and references therein).
For magnetic fields close to or exceeding the quantum critical
field, the contributions of the vacuum to the dielectric and
magnetic permeability tensors become important. In particular
when the plasma and vacuum contributions balance, a vacuum
resonance appears, where mode conversion may occur
\citep{Gnedin78,Pavlov79a,Meszaros79}.

For $B \leq B_q = 4.4  \times 10^{13}$~G, the vacuum parameter
is given by  (\citealt{Gnedin78}; see also
\citealt{Kaminker82})
\begin{equation}
W = \left( \frac{3\times10^{28} \rm{cm}^{-3}}{n_e} \right)
\left(\frac{B}{B_q}\right)^4
\end{equation}
where $n_e$ is the electron number density.

The ellipticities of the normal modes for  electromagnetic waves in the
magnetized plasma are  obtained from \citep{Kaminker82}
\begin{equation}
q = \frac{\sin^2 \theta }{2 \cos \theta} \sqrt{u} \left(1 - W\frac{u-1}{u^2}\right),
\end{equation}
where $u=\omega_{_{c,e}}^2 / \omega^2$,  $\omega_{_{c,e}} = eB/m_ec$ is
the electron cyclotron frequency and $\theta$ is the angle between
the wave propagation direction and the magnetic field.
The condition $q=0$ determines the frequencies and photon directions (for 
each value of density and magnetic field) at which
the normal modes approximation breaks downs (due to vacuum corrections
and proton resonance).
At this MCP a proper treatment of  radiative transfer
should be performed via the Stokes parameters, while our equations
are written assuming that the normal mode approximation  holds for
all photon propagation directions.
Following \cite{Pavlov79a}, we estimate the critical angle associated to
the normal modes breakdown  at the vacuum MCP from the expression
 \begin{equation}
\frac{\cos \theta_n}{\sin^2 \theta_n} = \pm \frac{1}{2\sqrt{u}}
\left[ \gamma_\parallel (u -1) + \gamma_\bot\left(1 - \frac{2W}{u}
\right)\right]_{q=0},
 \end{equation}
where $\gamma_{\parallel,\bot} = (\nu_{\parallel,\bot}  + \nu_r ) / \omega \sim 10^{-5} - 10^{-3}$. 
Here,  $\nu_r$ is a radiative width  and   $\nu_{\parallel,\bot}$ 
 is an effective electron-ion collision frequency 
 (for details see \citealt{Pavlov79a}).
At $\theta_n$  the normal modes have maximum non-orthogonality
(i.e. maximum coincidence).
If $\theta_n \to \pi/2$, the interval of non-orthogonality collapses to a
point and, for almost all waves (i.e. all those with $\beta = | \pi/2 -
\theta
| > \beta_n = | \pi/2 - \theta_n |$), the polarization ellipses are
rotated
through $90\deg$ while crossing the resonance (mode switching). Vice
versa, if $\theta_n$ is small
the evolution of the electromagnetic waves across the resonance is not
well understood. In the numerical simulation, we then set complete mode
conversion at the vacuum resonance as long as $\theta_n> 50~\deg$.

\begin{figure}
\centering
\includegraphics[width=8.0cm,viewport=10 5 450 350,clip]{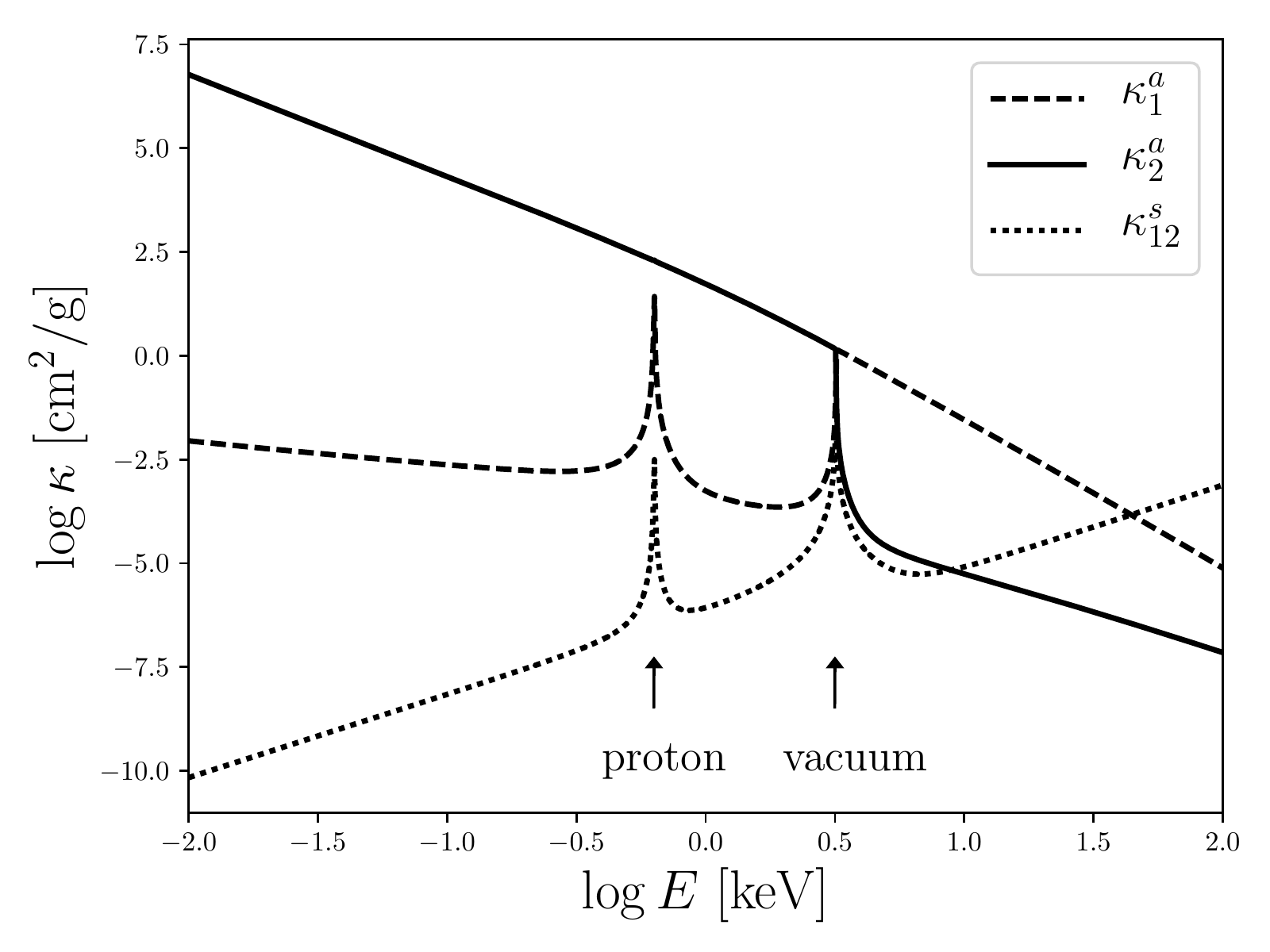}
\caption{Angle integrated opacities for a fully ionized H atmosphere
with magnetic field $B=10^{14}$~G (normal to the surface),  temperature
$T=10^7$~K and density  $\rho=10~\rm{gr~cm}^{-3}$. 
{  The dashed and 
solid lines correspond to the free-free absorption  opacities  in mode 
1 (X-mode) and mode 2 (0-mode), respectively.} 
 The dotted line correspond
to the scattering opacity from mode 1 to mode 2. Notice
that due to opacity  symmetry  $\kappa^s_{12} = \kappa^s_{21}$. The arrows
indicate the proton and vacuum  resonances.}
\label{fig:opacity}
\end{figure}

Notice that other important effects under strong magnetic field such 
as partial ionization  or the appearance
of additional narrow resonant features due to  higher  harmonics
of the proton cyclotron frequencies may be present (for a review see
\citealt{Potekhin14}).
However, for simplicity they are not considered in the present work, 
which is anyway restricted to a frequency integrated calculation.

Under the conditions we expect, conservative electron scattering
can be safely assumed. The  corresponding expressions for the opacity
are taken form \cite{Venture79} and \cite{Kaminker82}, while those for
ions  from  \cite{Pavlov95}.
Opacities for  both electrons and ions bremsstrahlung are from
\cite{Pavlov76}, \cite{Meszaros92} and \cite{Pavlov95},
and we implemented  the numerical evaluation of the magnetic Gaunt
factors as in  \cite{Zane00}.

\begin{figure*}
\includegraphics[width=8.0cm,viewport=5 0 450 350,clip]{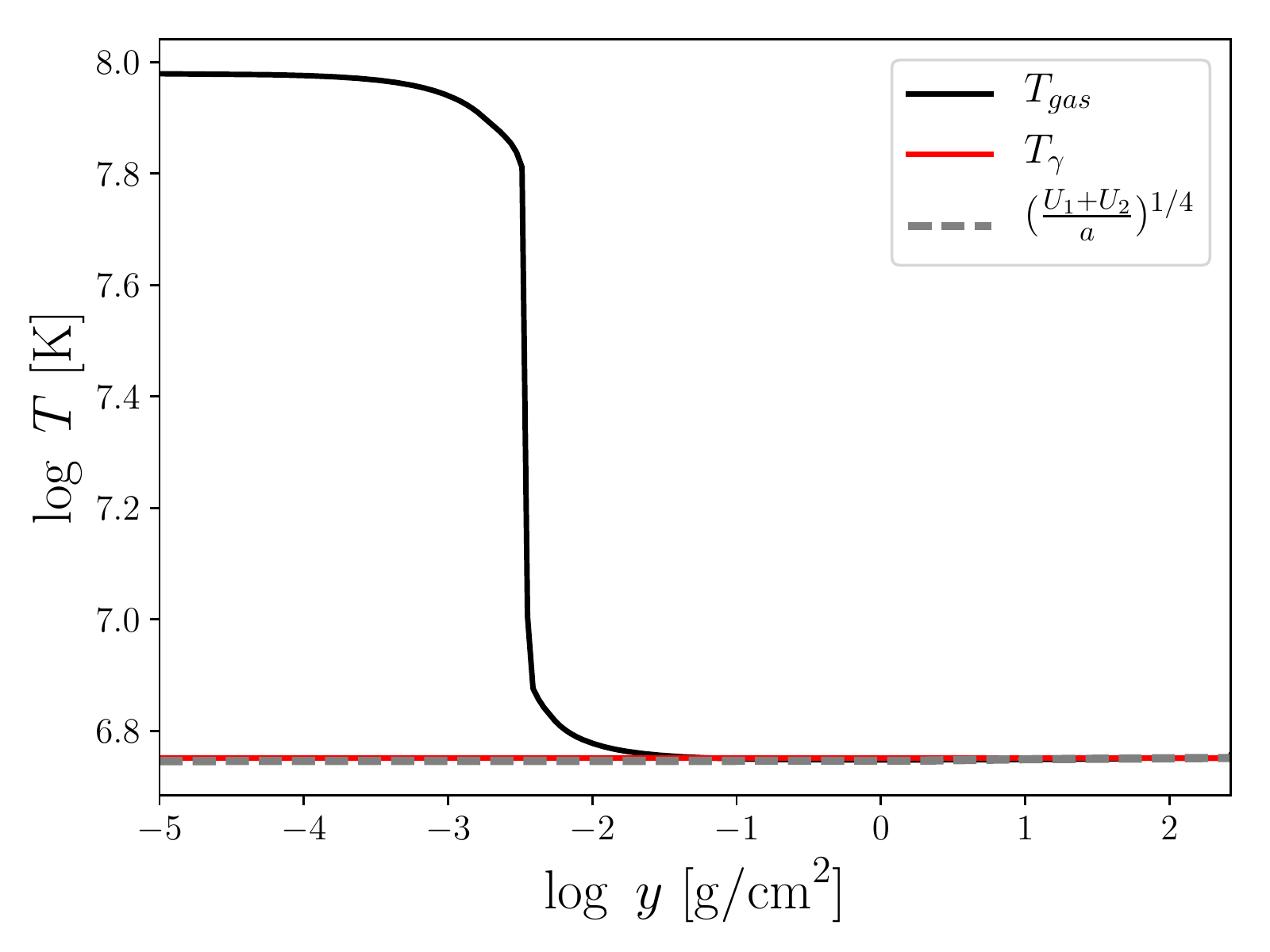}
 \includegraphics[width=8.0cm,viewport=5 0 450 350,clip]{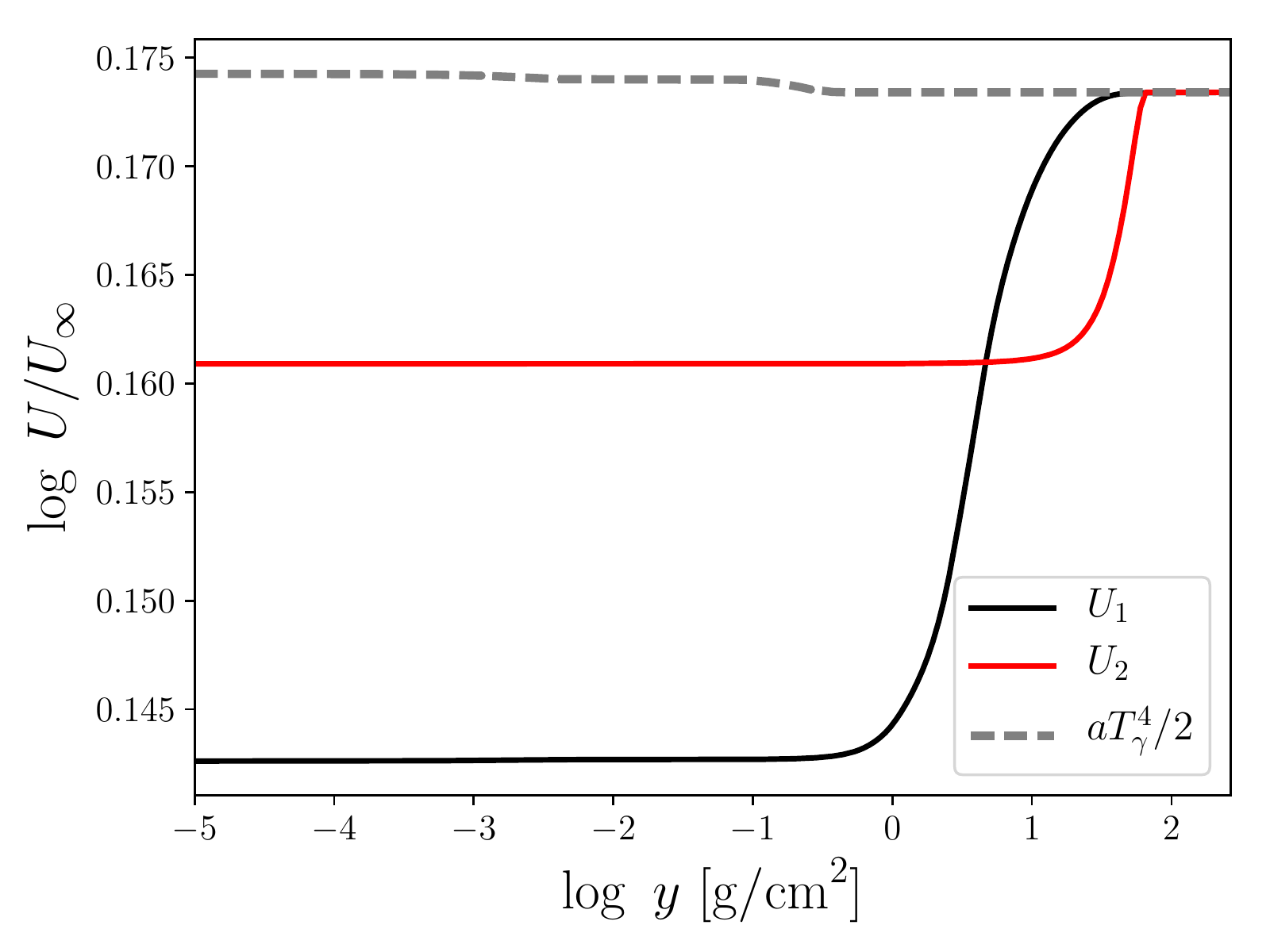}
\includegraphics[width=8.0cm,viewport=5 0 450 350,clip]{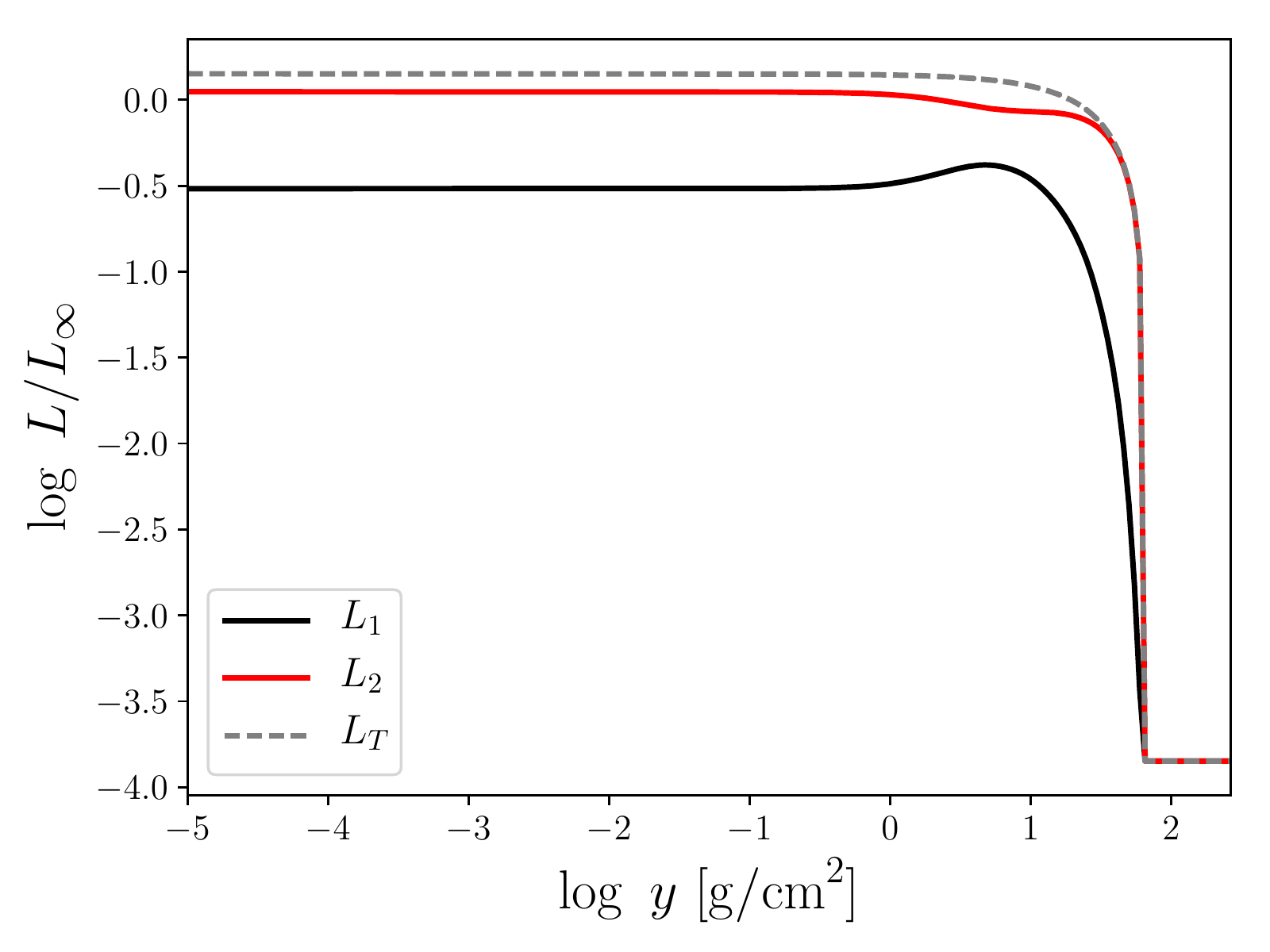}
\includegraphics[width=8.0cm,viewport=5 0 450 350,clip]{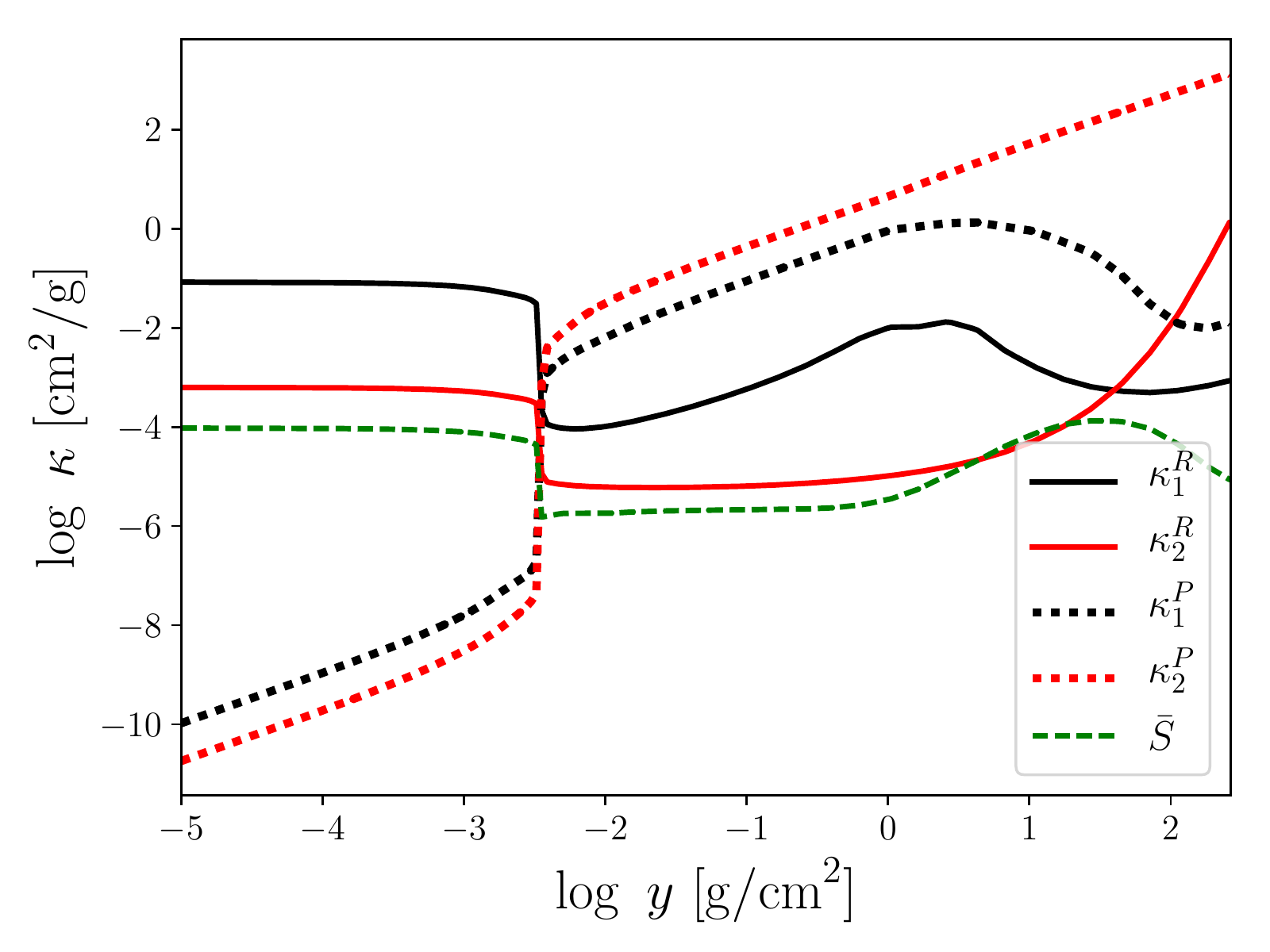}
\caption{Gray atmosphere solution for total luminosity $L_{\infty} = 10^{36}~\rm{erg/s}$,
stopping column density $y_{_0} = 65 ~ \rm{gr/cm}^2$
and magnetic field $B= 10^{14}~\rm{G}$. {\bf Top-left panel}:
gas temperature profile, $T$ (black solid line), and
photon temperature profile,  $T_\gamma$ (red solid line). The temperature  associated to the
total energy density is shown in the gray dashed  line.
{\bf Top-right panel}: energy densities associated to
 $U_1$ (black solid line), $U_2$ (red solid line) and $T_\gamma$ (dashed gray line)
 normalized to $U_\infty = L_\infty/ 4\pi R^2 c$. {\bf Bottom-left panel}:
 luminosity profiles $L_1$ (black solid line) and $L_2$ (red solid line).
 The total (analytical) luminosity $L_T$ is shown in the gray dashed line.
 {\bf Bottom-right panel}: Planck mean opacities,  $\kappa_1^{_P}$ (black dotted line) and
 $\kappa_2^{_P}$ (red dotted line), and  Rosseland mean opacities, $\kappa_1^{_R}$ (black
 solid line)  and $\kappa_2^{_R}$ (red solid line). The scattering mean opacity from mode 1
 to mode 2 (and vice versa), $\bar S$,  is shown in the green dashed line.
}
\label{fig:first}
\end{figure*}

Figure \ref{fig:opacity} shows the energy dependent  free-free absorption and the
mode exchange scattering opacity  used in our calculations. In general, for most
energies, opacities are dominated by photon-electron interaction. However, for the magnetic
fields that we are considering, $B\sim 10^{14}$~G, protons also play an
important role as they produce
a sharp feature in the X-mode opacities at the proton cyclotron energy, which lies
within the soft X-ray band (see \citealt{Zane01}, \citealt{Ho01} and
\citealt{Ozel03}
for the discussion of the impact of this feature in the emergent spectrum from
atmospheres of magnetized NSs). On the other hand, vacuum effects introduce additional
features to the opacity profiles. In fact, when complete mode conversion is present,
vacuum effects produce a step change in the free-free opacities across the
vacuum resonance, due to the exchange of the X-mode opacity
with the O-mode opacity
(and vice versa;
see \citealt{Ho03a} for a discussion of the vacuum resonance with and without mode conversion).
Instead, vacuum effects produces a raise (spike feature) in
the (angle-integrated) mode exchange scattering opacity at the vacuum
resonance.

As mentioned in \S~\ref{radiative}, in our numerical
calculation we use the above opacities suitably integrated in frequency in
strict analogy with the definition of
the absorption mean opacity and the flux mean
opacity in the unmagnetized case. A further problem encountered while
performing the integration in energy is related to the fact that the
vacuum resonance is located at different energies in different layers of
the atmosphere.
To avoid overestimating the mean opacity due to the finite
size of the energy grid, which can lead to numerical problems of
convergence in
our calculations, we compute the opacity near the vacuum resonance feature
at the
energy point $E_v$ as an average between its values at two
adjacent energies $E_{-} = E_v  -
\Delta E_{bin}/3$ and $E_{+} = E_v + \Delta E_{bin}/3$. We tested
numerically that varying the way in which we chosen the adjacent energies
only produces negligible differences in the resulting integral.

\begin{figure*}
\includegraphics[width=8.0cm,viewport=5 0 450 350,clip]{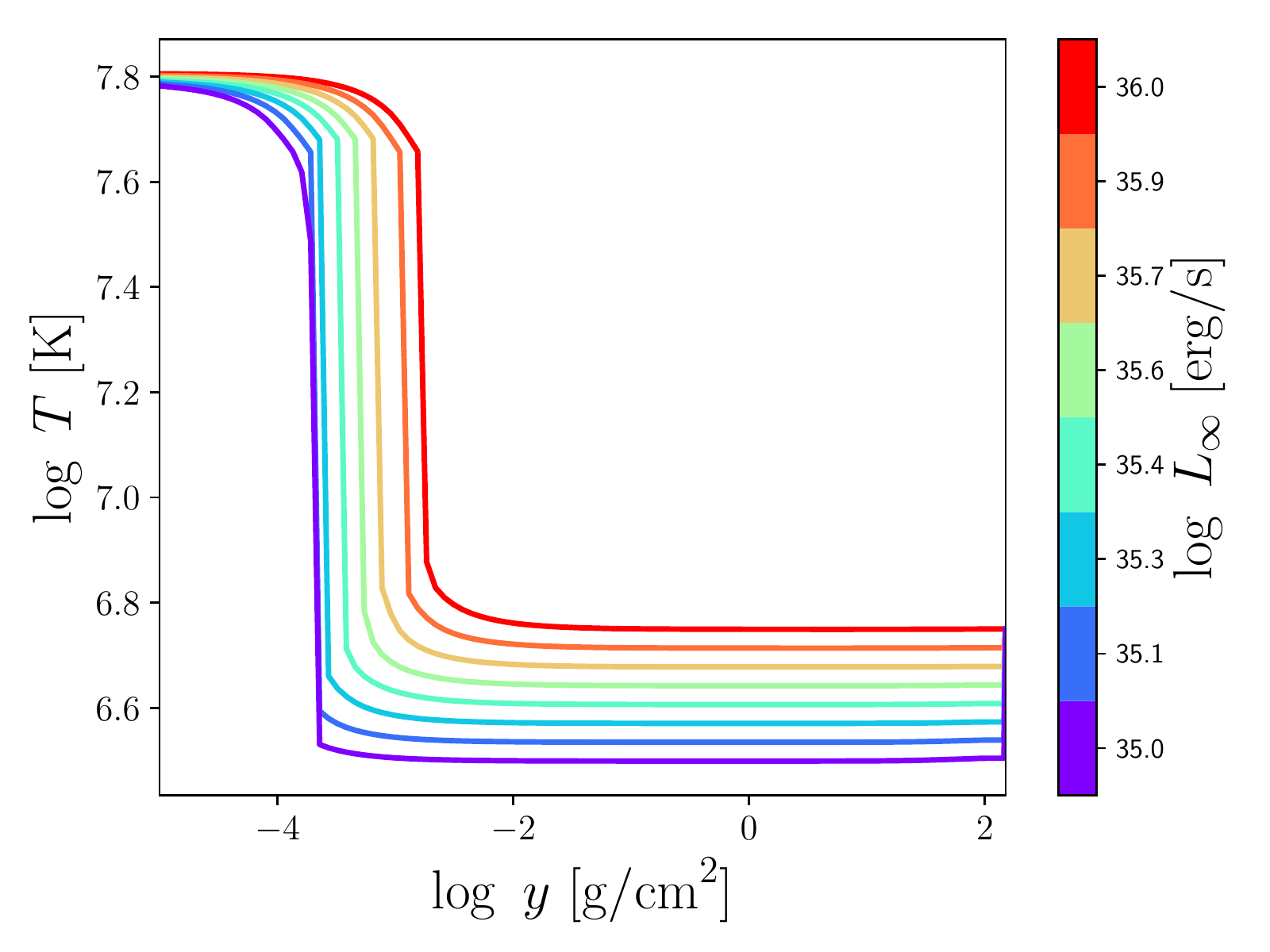}
\includegraphics[width=8.0cm,viewport=5 0 450 350,clip]{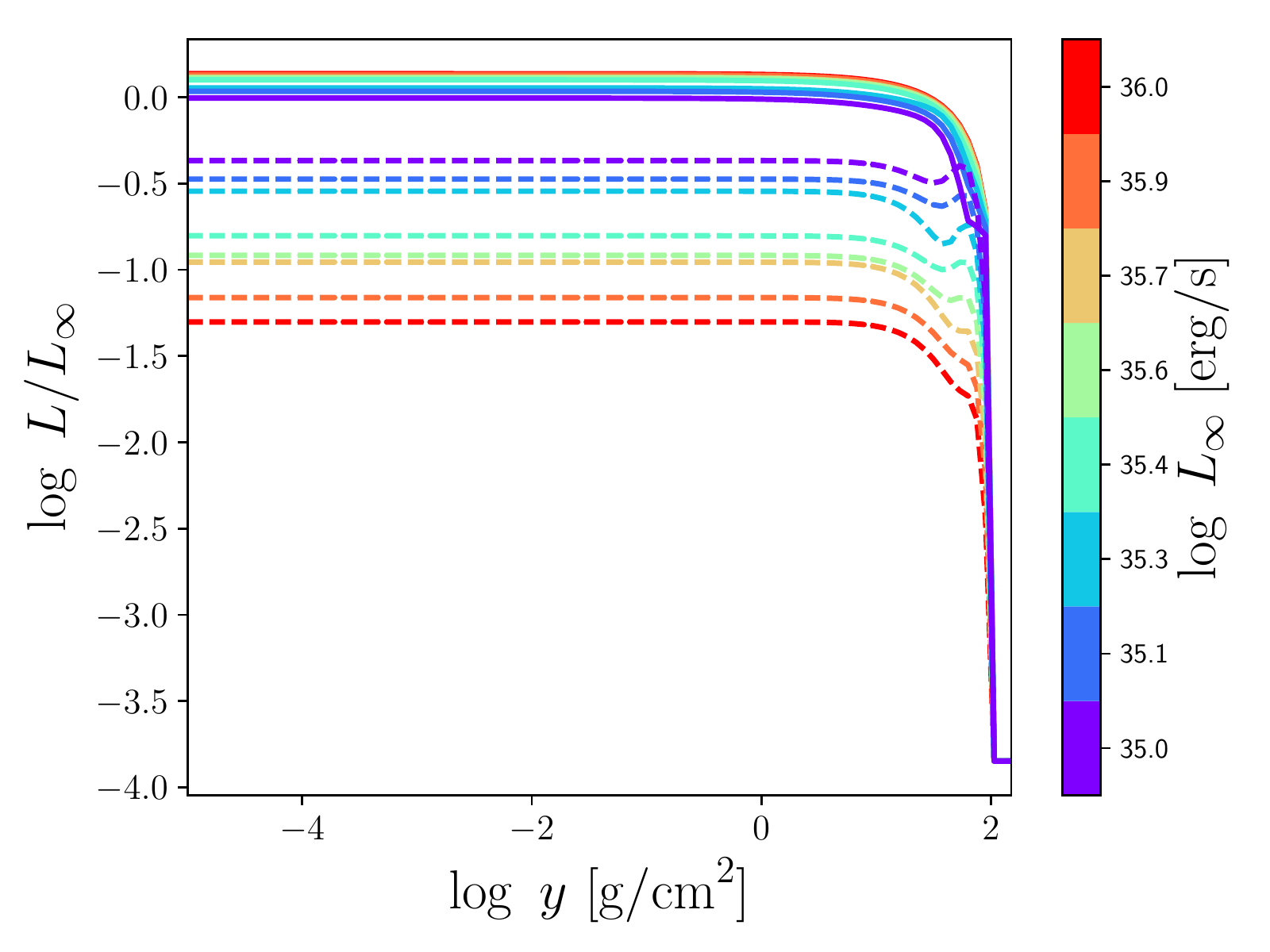}
\includegraphics[width=8.0cm,viewport=5 0 450 350,clip]{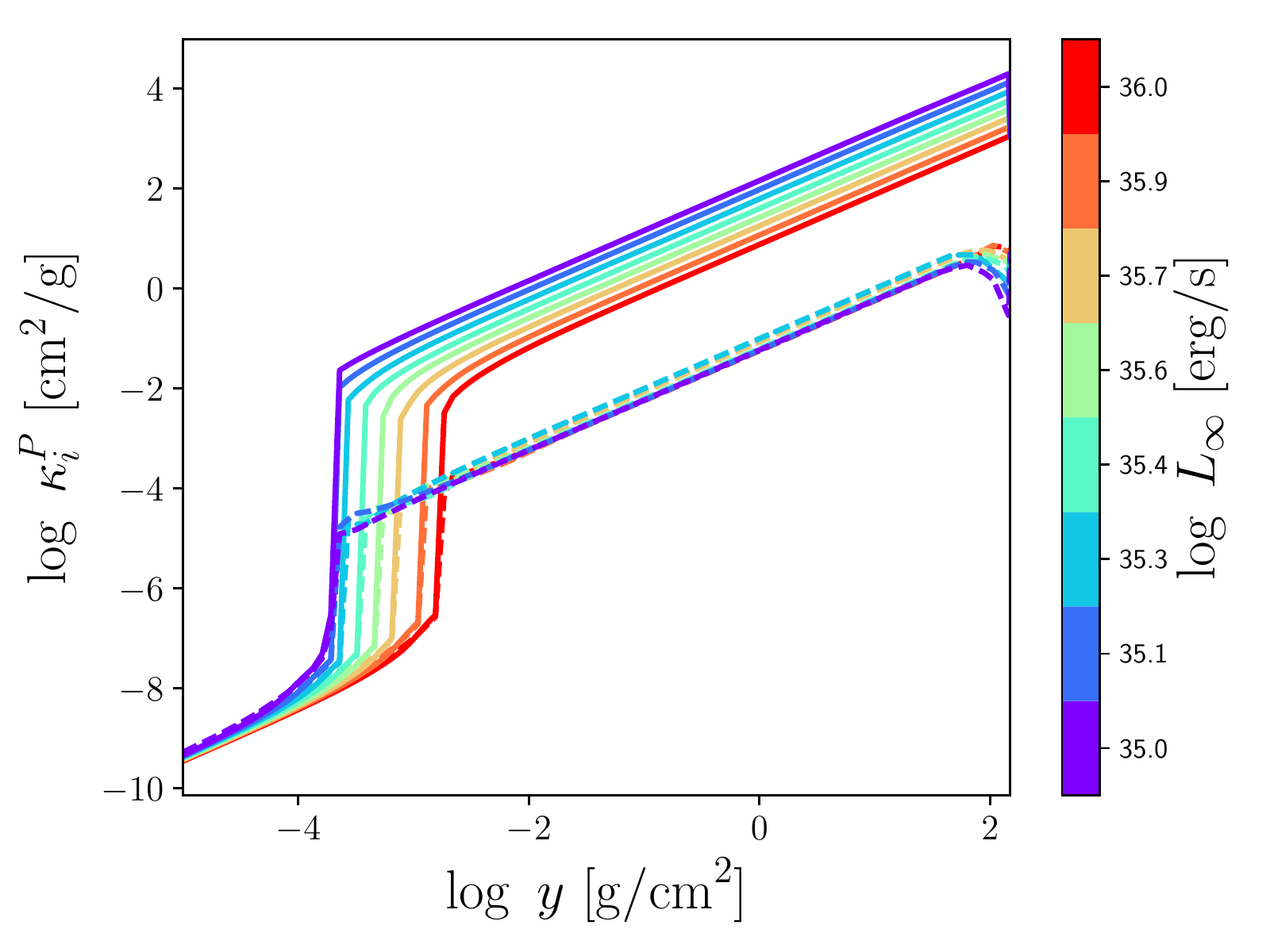}
\includegraphics[width=8.0cm,viewport=5 0 450 350,clip]{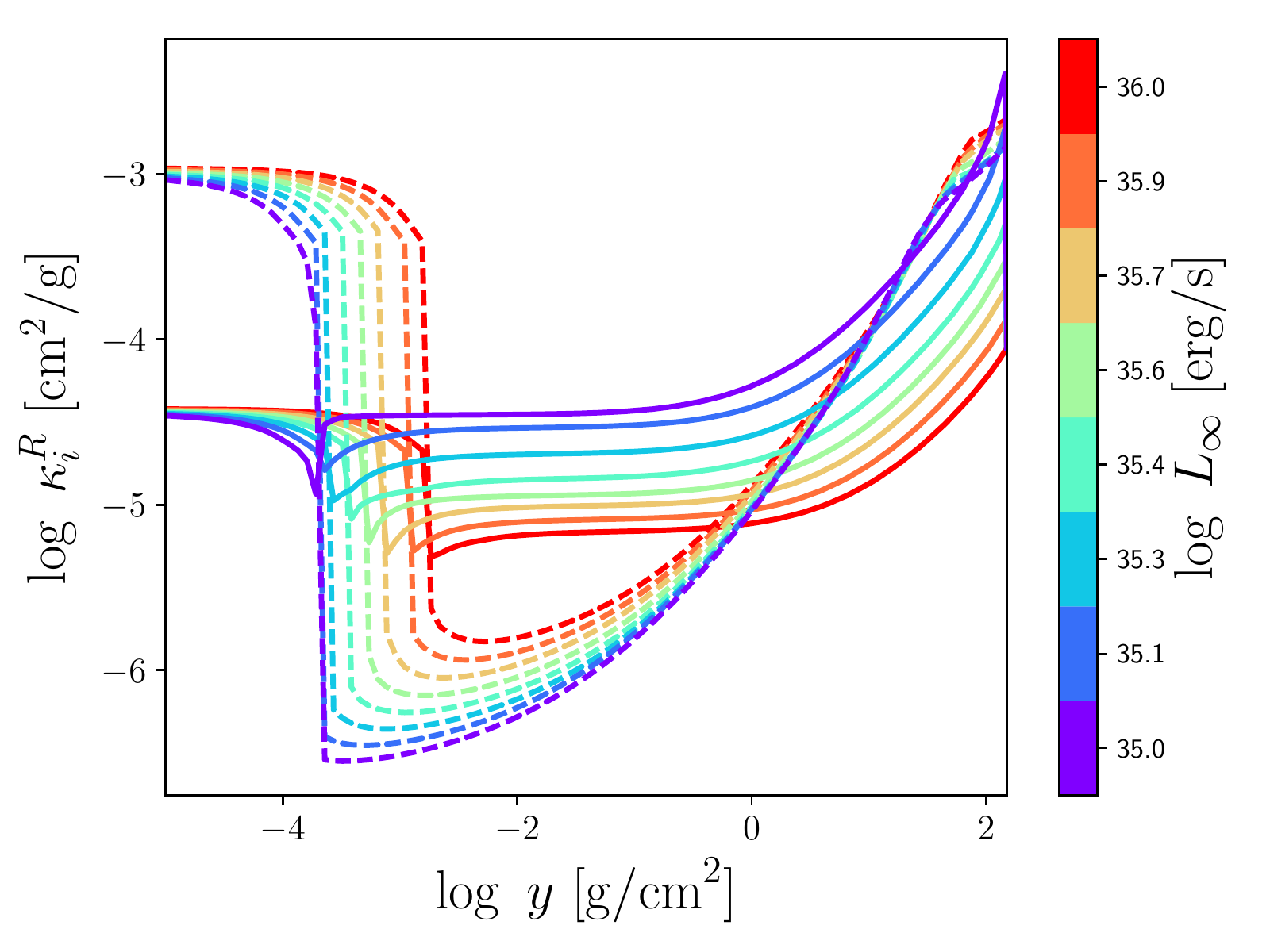}
\caption{Gray atmosphere solutions for  different luminosities
(see vertical  color bar), fixed magnetic field $B = 3\times 10^{14}$~G,
and fixed stopping  column density $y_{_0} = 100~\rm{gr~cm}^{-2}$.
The temperature profile for the atmospheric plasma is shown
in the top-left panel. The luminosity $L_1$ (top-right panel),
Planck  mean opacity $\kappa_1^{_P}$ (bottom-left panel),
and Rosseland mean opacity $\kappa_1^{_R}$ (bottom right panel)
are shown by dashed lines. Similarly, the corresponding profiles
for $L_2$, $\kappa_2^{_P}$, and $\kappa_2^{_R}$ are shown by solid
lines.}
\label{fig:seq_L}
\end{figure*}

\section{Numerical solutions}
\label{numsol}
{ The numerical calculation was performed by readapting the
code presented by \citet[][]{Turolla94}, which was meant for
unmagnetized atmospheres under accretion, to the case of magnetized NS 
atmospheres heated by particle bombardment. We started from the numerical 
routines for the opacities from \citet{Zane00} and created all necessary routines
 for the computation of the energy integrated opacities required by this work (see previous section).
Solutions are then found using a numerical iterative scheme.}
First, an initial temperature profile  is
specified, taking as trial  solution the temperature profile of
a non-magnetized
atmosphere from \cite{Turolla94}. This, together with
eq.~(\ref{eq:density})
sets also the initial density profile.
Then, we solve the transport of
radiation via  eqs. (\ref{eq:transfer1}), (\ref{eq:transfer2}) and  (\ref{eq:compton})
to obtain the profiles for $L_i$, $U_i$ and $T_{\gamma}$. These, in turn,
are used to compute a new temperature profile, that satisfies the
energy balance, eq.~(\ref{eq:balance}).
The procedure is iterated until the relative difference between two
successive solutions is  $<10^{-5}$, in each variable. Convergence
is typically reached after $\sim 15$ iterations.

In all models, a NS with mass $M= 1 M_\odot$ and
radius $R = 10$~km is assumed\footnote{If we vary the mass or
radius by a factor 2, our numerical solutions change less than $2\%$.}. 
Solutions are computed by using a logarithmic grid
of 100 equally spaced  points for the column density
$y$ in the range $[10^{-5}-10^{2}]~\rm{gr~cm}^{-2}$.
We report the models for magnetic field values
$B\approx 10^{14}-10^{15}$~G and
luminosities  $L_\infty \approx 10^{35} - 10^{36}~\rm{erg~s}^{-1}$, which
are
characteristics of the brightest magnetars, and stopping lengths in the
range  $y_0 = 65 - 500~\rm{gr~cm}^{-2}$ (see \S~\ref{stopping}).

\subsection{Results}
\label{res}

Fig.~\ref{fig:first} shows the temperature, energy density, luminosity
and opacity profiles for a NS atmosphere with
$B= 10^{14}$~G, stopping length $y_0=65~\rm{gr~cm}^{-2}$ and total
luminosity $L_\infty = 10^{36}~\rm{erg~s}^{-1}$.
The temperature profile (top-left panel) shows the formation
of a hot layer in the most external part of the atmosphere.
This external temperature profile  is similar to those  obtained in the case
of atmospheres of NS accreting at low rate from the interstellar medium
(see \citealt{Turolla94} and  \citealt{Zane00}).
As discussed in previous works, free-free cooling
is  not effective in balancing the heating
(in this case caused by the particle bombardment) in this region. As a
consequence, the plasma temperature raises until a point in which
Compton cooling becomes efficient (see \citealt{Zane00}
and references therein). Since the  radiation is produced
only by the
particle bombardment (or, alternatively, the luminosity by particle
bombardment is much larger than the NS cooling luminosity), in the
deepest layers the luminosity vanishes and the temperature profile becomes
constant.
The radiation temperature shows that  most  of the  radiation decouples from
the plasma in the  internal region  of the atmosphere, before photons
start travelling across  the hot external layer.

\begin{figure*}
\includegraphics[width=8.0cm,viewport=5 0 450 350,clip]{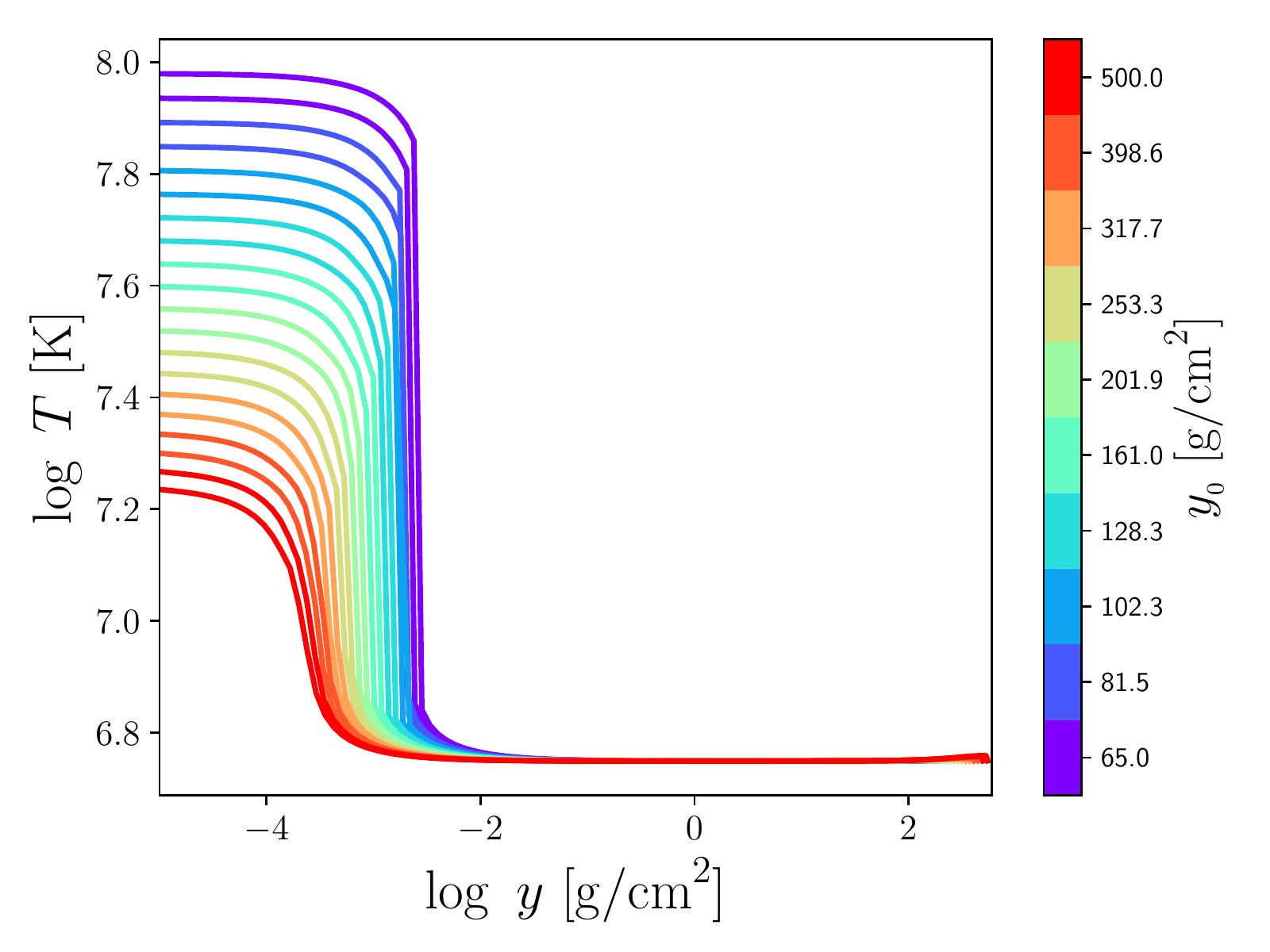}
\includegraphics[width=8.0cm,viewport=5 0 450 350,clip]{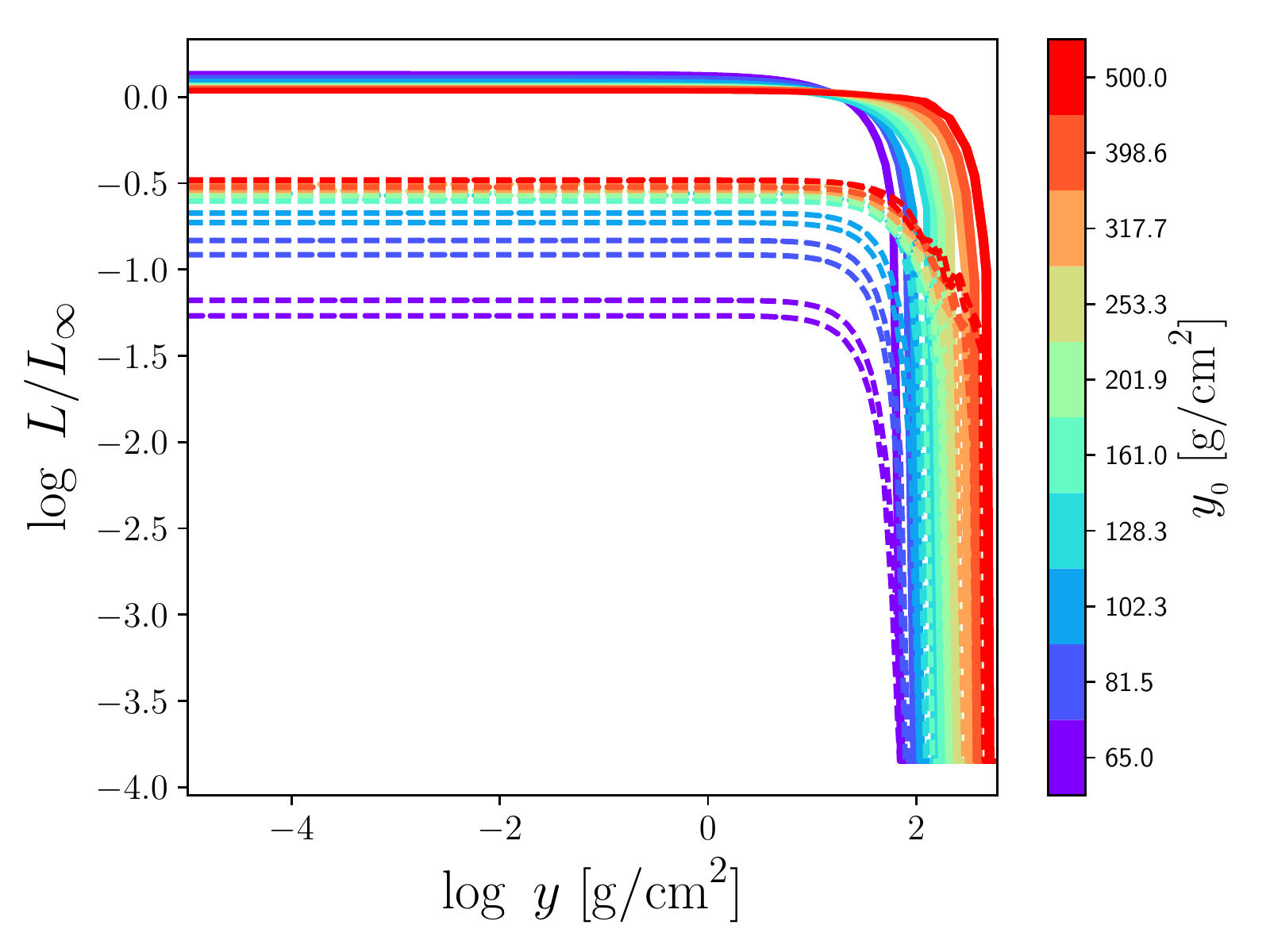}
\includegraphics[width=8.0cm,viewport=5 0 450 350,clip]{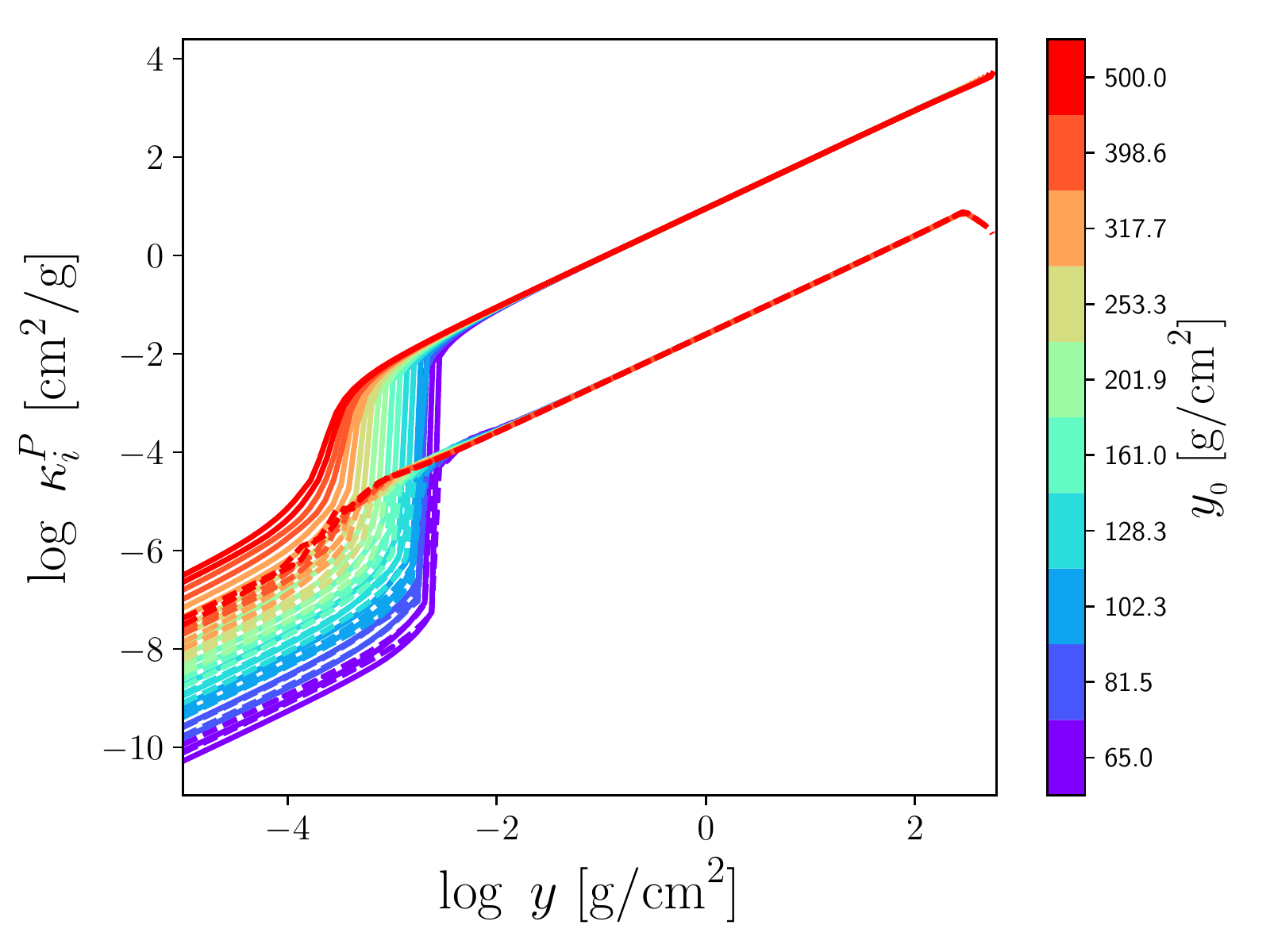}
\includegraphics[width=8.0cm,viewport=5 0 450 350,clip]{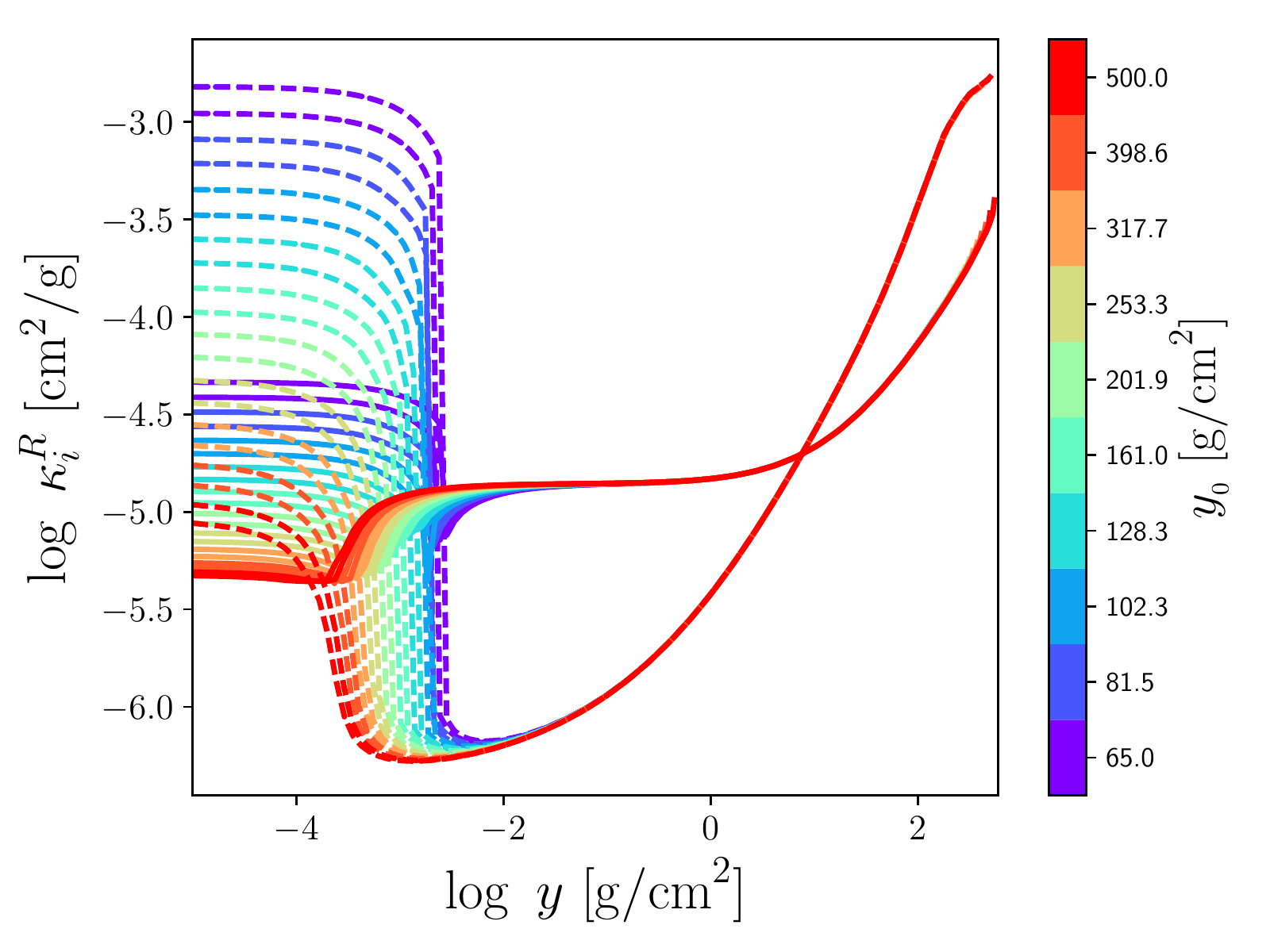}
\caption{Same as Fig. \ref{fig:seq_L}, but varying the stopping column
density $y_{_0}$. Here,  the total luminosity is set
to $L_\infty = 10^{36}~\rm{erg/s}$ and the magnetic field to
$B=4\times 10^{14}$~G.}
\label{fig:seq_Y}
\end{figure*}

The top-right and bottom-left panels of Fig. \ref{fig:first} 
show the energy density and the luminosity
of the two modes. Clearly, $U_i$ is constant in the internal layers where
the luminosity vanishes and energy equipartition is reached between the
two radiation modes. As matter and radiation start decoupling, most of
the luminosity is produced in an internal region, characterized by a
substantial and non
vanishing energy density gradient, $\der U_i/\der y$, in which
diffusion approximation holds. It is interesting
to notice how the energy density profiles cross, at some point within the
atmosphere,
reflecting the mode conversion
due to vacuum effects (Fig. \ref{fig:first}, bottom-right panel).
This is an interesting effect since it can impact on the
expectations for the polarization signal. In fact, since the plasma is highly
magnetized, the emerging radiation is expected to be highly polarized, in
principle even up to 100$\%$: the
large difference in the plasma opacities of the two modes implies that
most of the
radiation should be carried by the mode for which the atmosphere is more
transparent (see for example  \citealt{Ozel01} and \citealt{Ho03a}
for a discussion in  the context of atmosphere for
passively cooling NSs).
However, vacuum mode conversion partly reduces the difference
in the opacities, and therefore in luminosity, of the two modes,
potentially reducing the polarization signal. If an
``intrinsic'' polarization fraction is defined as
$PF = (L_{\infty,1} - L_{\infty,2})/(L_{\infty,1} + L_{\infty,2})$,
then $PF \approx 0.6$ in the model at hand.

In the following, we explore the numerical solutions for different
values of luminosity, stopping length and
magnetic field. In order to isolate the effects of each parameter, we
vary one of these quantities at each time, keeping fixed the other
model parameters.
We present the profiles for temperature, luminosity, Planck
and Rosseland mean opacities as they are the most relevant to discuss the
behaviour of the solution (the scattering opacity is in fact negligible in
most of the atmosphere).

\subsubsection{Models with different luminosity}
\label{lum}

Fig. \ref{fig:seq_L} shows a set of  solutions computed
for $y_{_0}=100~\rm{gr~cm}^{-2}$, $B=3\times 10^{14}$~G, and
by varying the total luminosity in the range $L_\infty
=10^{35}-10^{36}~\rm{erg~s}^{-1}$.  As expected, higher luminosities, which
are related to larger energy deposition in the atmosphere,
correspond to hotter atmospheres. Moreover, by increasing the luminosity,
the hot atmospheric
layer extends inwards, to higher column densities.
Luminosity and temperature profiles
scale in a roughly self--similar way.
However, the expected emergent polarization increases for
higher luminosities. This can be understood noticing that, in the inner
regions where most of the luminosity is created,
the difference between the Rosseland mean opacities
associated to each normal mode (lower right
panel), is larger for higher luminosities. In diffusion approximation,
this translates in a larger suppression of the radiation  carried by a
mode with respect to the other, and the difference in luminosity then
propagates till the top of the atmosphere.
For the models presented here, the resulting emergent polarization varies
between $0.4 < PF < 0.9$ for the range of luminosities $L_\infty$
considered.

{
The behaviour for the opacities in these models can be understood as follows. 
The Planck mean opacity, $\kappa^{_P}_i$, is obtained by performing an
integration over energies of the free-free absorption opacity weighted by a blackbody. In turn, the integrand over
  energy is dominated by the contribution of the free-free opacities
    at relatively  low energies, which are less sensitive to the presence
   of the vacuum resonance and   subsequent mode switching. It follows that, for a fixed magnetic field, 
   the integrand in the   opacity scales approximatively as $\kappa^{\rm{ff}}_i \propto y T^{-7/2} f_i(E)$, where $f_i(E)$ 
 is a function of the energy. Since most of the atmosphere has a uniform temperature, this translates
into a Planck mean opacity with a linear dependency on $y$. In the external hot layer the temperature
 increases, and this produces the observed drop in the Planck mean opacities according to 
 the $T^{-7/2}$ dependency of the free-free opacity. 
 
 The Rosseland mean opacities, instead, are computed considering the contribution of both,
  the free-free and scattering opacities. In particular, scattering opacities are dominant at low 
  densities, which is reflected on fact that, in the most external atmospheric layers, the Rosseland 
  mean opacity tends toward a constant. Instead, the free-free opacity is dominant at high densities. 
  Moreover, since in the Rosseland mean opacity the associated integration
   is performed over the inverse of the energy-dependent opacity, 
  the major contribution from free-free opacities is from the higher energies which are more sensitive to the vacuum 
  resonance. This in turn introduces an additional density dependency, which is reflected in the 
  non linear growth of the Rosseland means at large $y$.
}

\begin{figure*}
\includegraphics[width=8.0cm,viewport=5 0 450 350,clip]{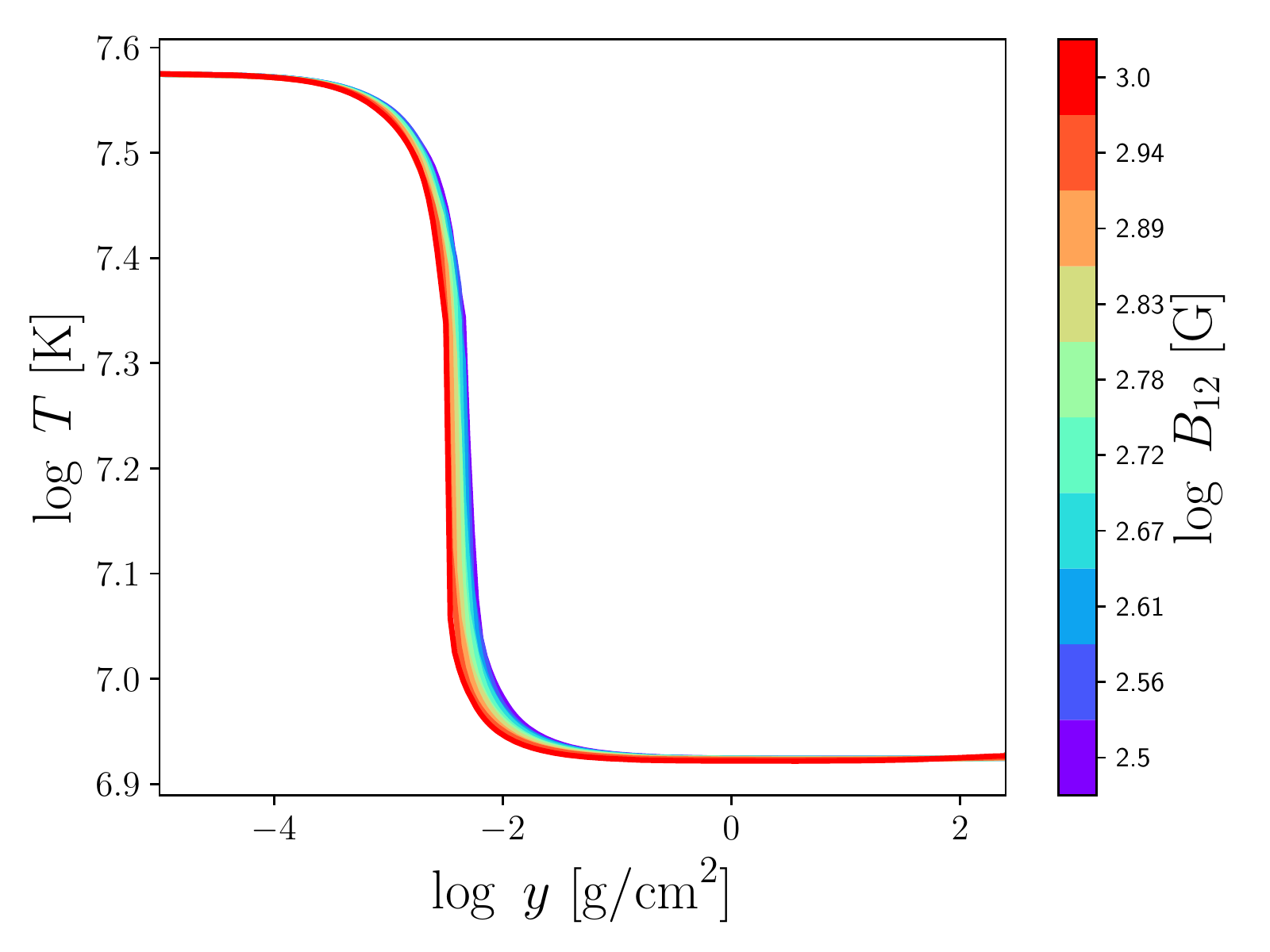}
\includegraphics[width=8.0cm,viewport=5 0 450 350,clip]{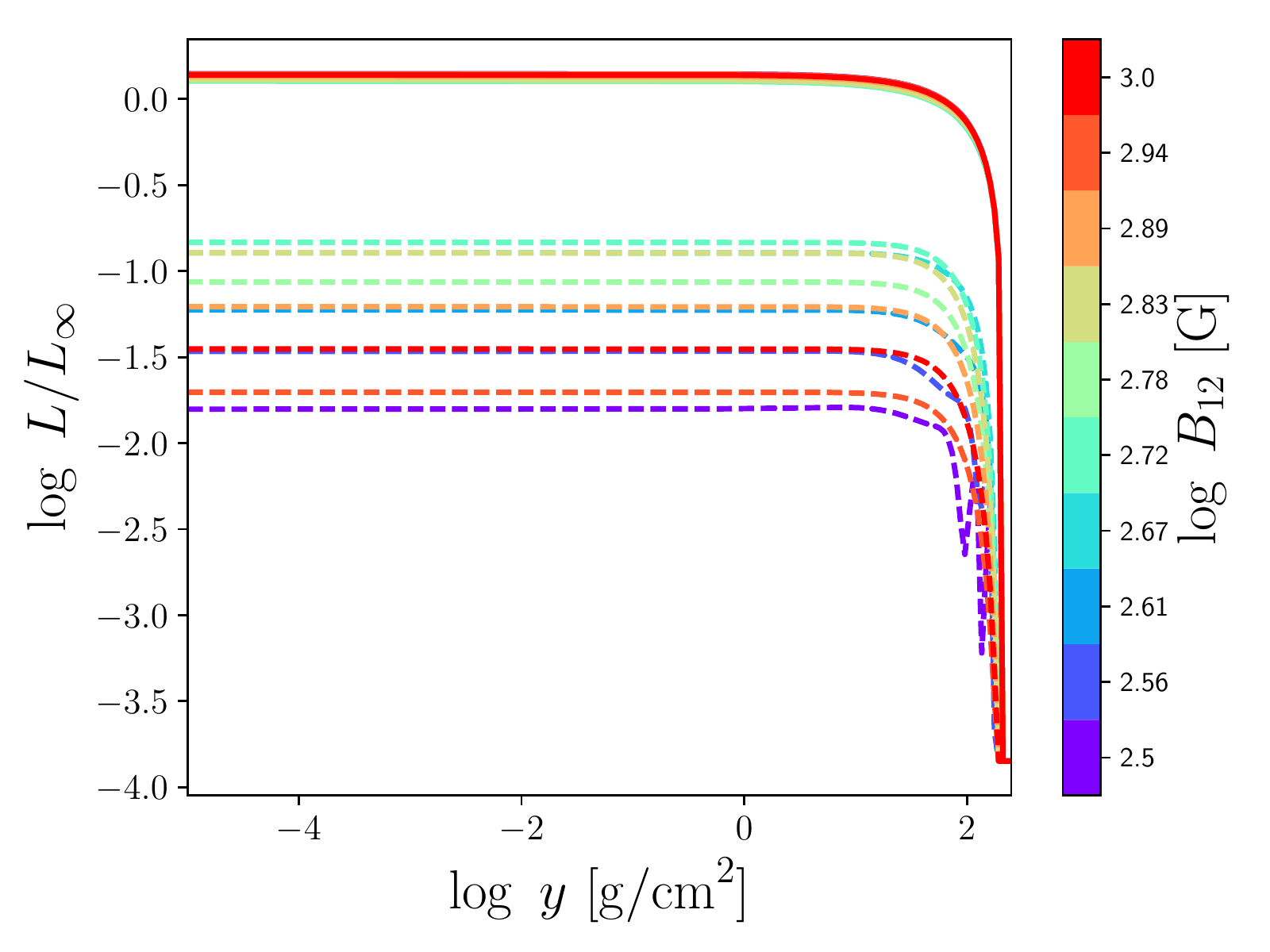}
\includegraphics[width=8.0cm,viewport=5 0 450 350,clip]{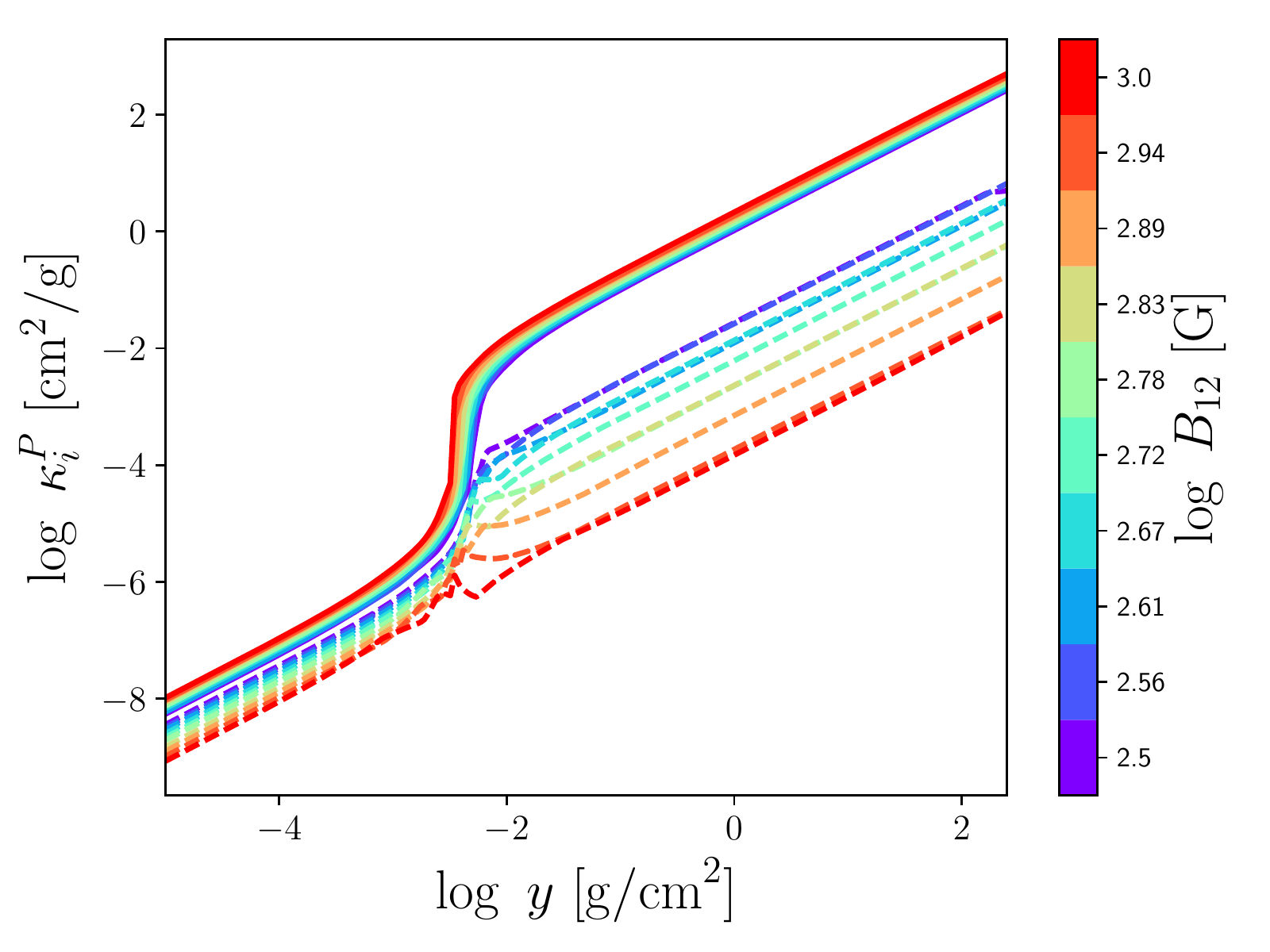}
\includegraphics[width=8.0cm,viewport=5 0 450 350,clip]{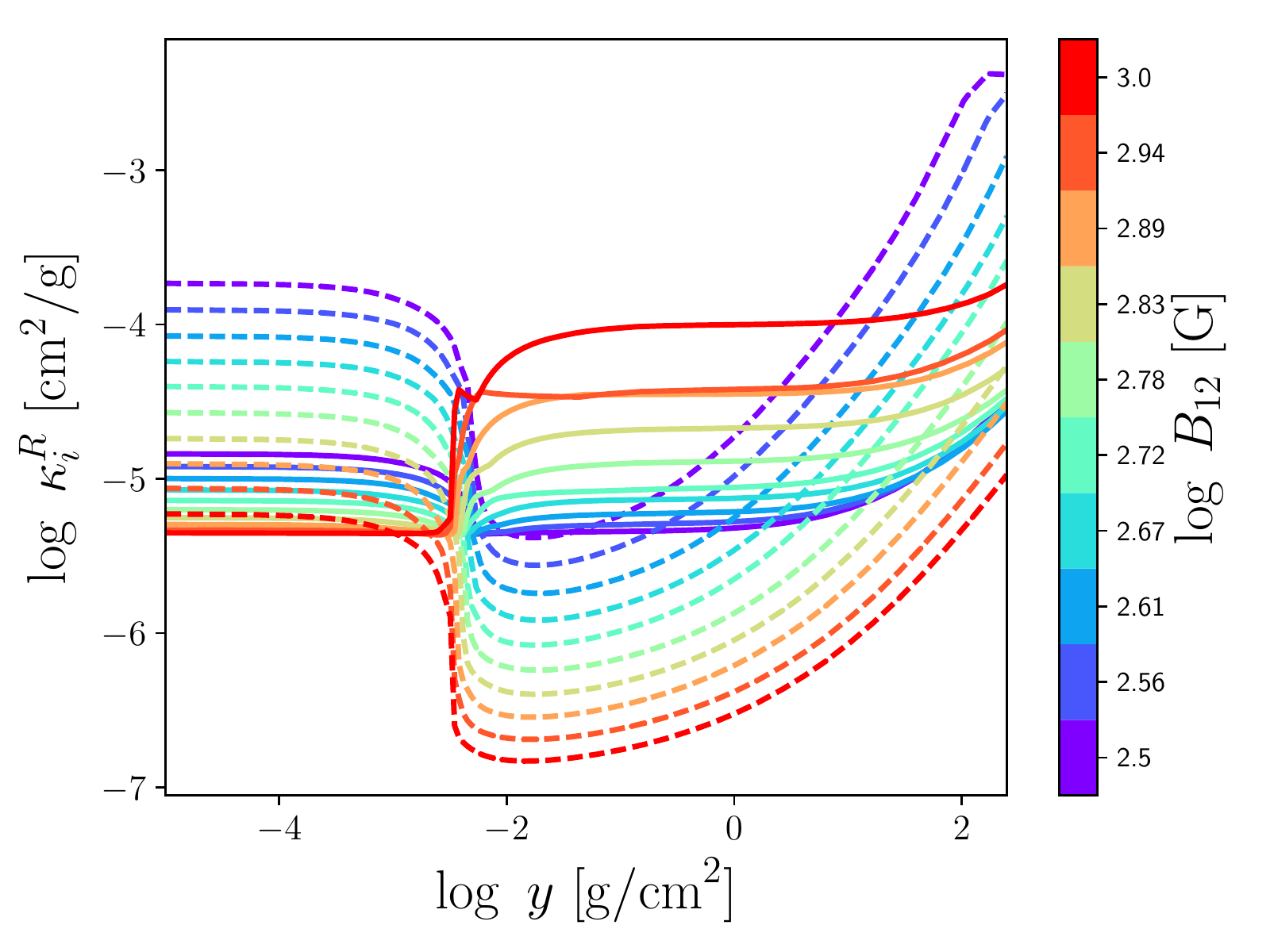}
\caption{
Same as Fig. \ref{fig:seq_L}, but varying the magnetic field. Here,
the total luminosity is set to  $L_\infty = 5\times 10^{36}~\rm{erg/s}$ and the
stopping column density to $y_{_0}=200~\rm{gr/cm}^2$.}
\label{fig:seq_B}
\end{figure*}

\subsubsection{Models with different stopping column density}
\label{colden}

Fig. \ref{fig:seq_Y}  shows the solutions   for stopping column
densities in the range $y_{_0} = 50 - 500~\rm{gr~cm}^{-2}$,
 total luminosity $L_\infty = 10^{36}~\rm{erg~s}^{-1}$, and  magnetic
field $B=4\times 10^{14}$~G.
At variance with the models in Sec. \ref{lum}, the stopping column density  $y_{_0}$
is anticorrelated  with temperature of the  hot external layer.
Indeed, for smaller stopping column densities, the energy
deposition is distributed in a smaller region (see eq.
\ref{stopping}), which leads to external  layers  with  higher
temperatures, as shown in the top-left panel.
In this case, the hot layer moves inwards as $y_{_0}$
decreases. In addition, the most internal layers of the atmosphere have the
same temperature for different $y_{_0}$. This is  interesting
for future frequency-dependent calculations since it
indicates that the typical photon energy
deep in the atmosphere is quite
 insensitive  to variations of $y_{_0}$. Therefore,
 the spectrum produced in the atmosphere interior
 may be   determined only by the   total luminosity and magnetic
 field strength.

The luminosity carried by  each normal mode is shown in the right upper panel
of Fig. \ref{fig:seq_Y}. Similar to the solutions for  the temperature profile,
the  polarization is  also anticorrelated with the stopping column density.
At variance with models presented in Sec. \ref{lum}, the region where
the  difference in the Rosseland  mean opacities of the two modes,   $\kappa_1^{_R}$ and $\kappa_2^{_R}$,
leads to a  larger polarization, is now located in the most external layers
of the atmosphere.

{

As we can see from the left top panel, in the models presented in this section 
the only change between the solutions is in the temperature (and location) of
 the external hot layer, while the density profile and the magnetic field are the same 
 for the different models. Accordingly, the strong dependency on $T^{-7/2}$ of 
 the free-free absorption introduces the drop and dispersion between the various
  curves of the Planck mean opacities seen in the most external atmospheric layers.

The corresponding dispersion in the values of the Rosseland mean opacities at
 low $y$ (scattering dominated) is instead due to the fact that they
 are evaluated by convolving the energy-dependent opacity over the derivative of the blackbody,
  $\mathrm{d} B_E(T)/\mathrm{d}T$, calculated at the local value of $T$. For higher
   temperatures, the peak of this function is located at higher energies. In other words, 
   if the photon bath in the external layer is hotter, then the major contribution to the 
   averaged Rosseland opacity is due to scattering of photons with higher frequency
    (which in turn may also be more largely affected by vacuum effects). 
 }

\begin{figure*}
\includegraphics[width=8.0cm,viewport=5 0 450 350,clip]{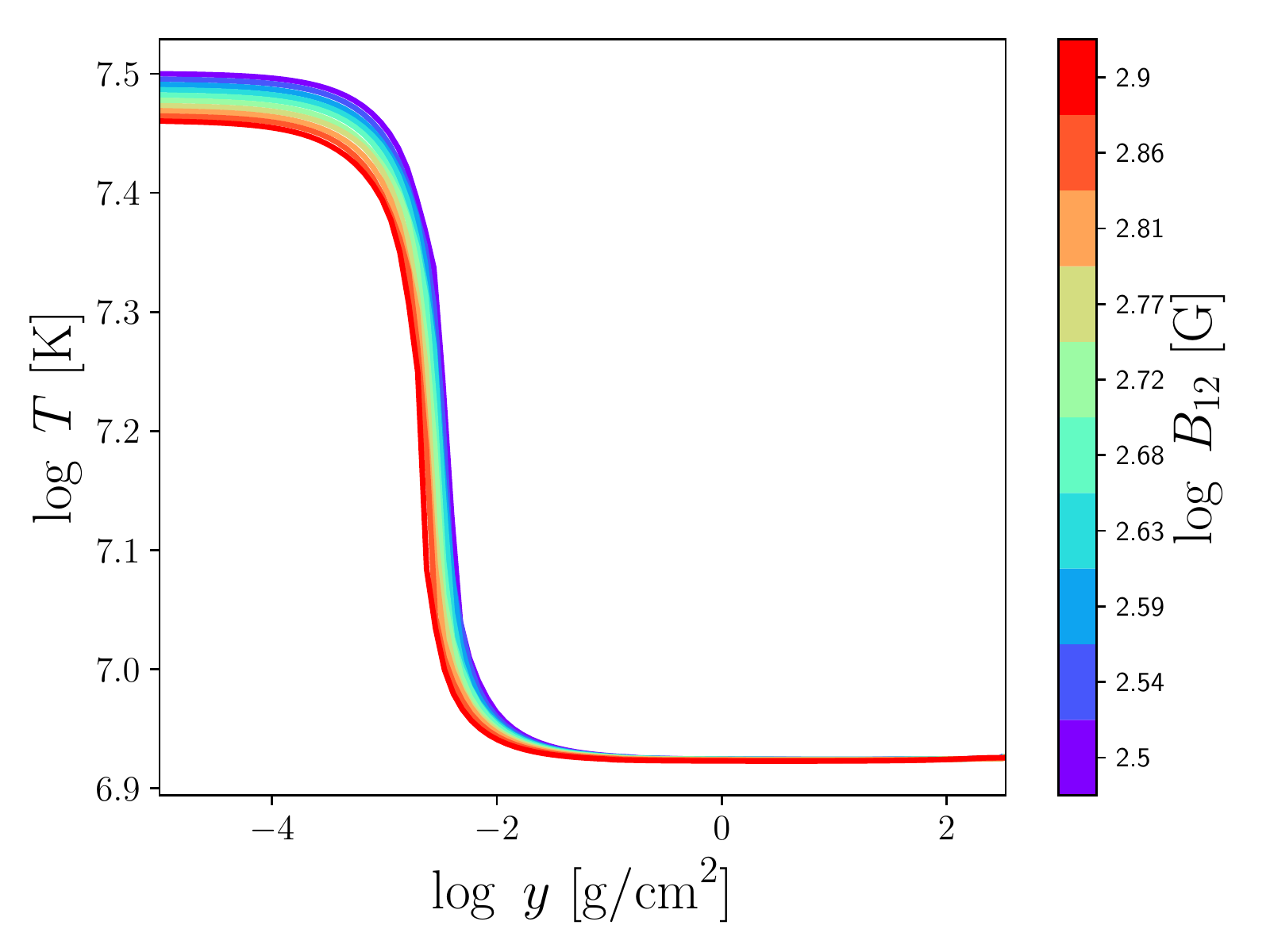}
\includegraphics[width=8.0cm,viewport=5 0 450 350,clip]{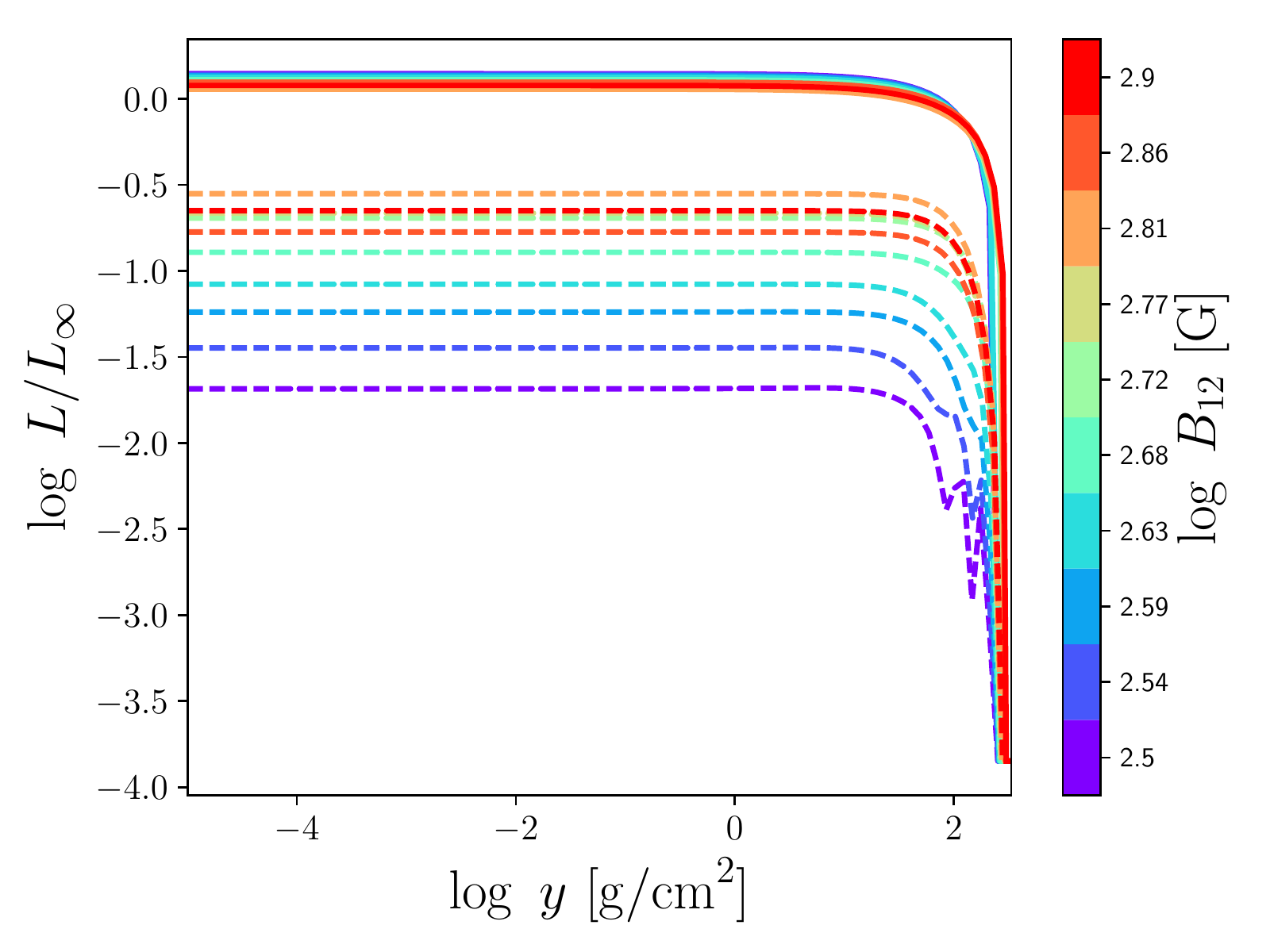}
\includegraphics[width=8.0cm,viewport=5 0 450 350,clip]{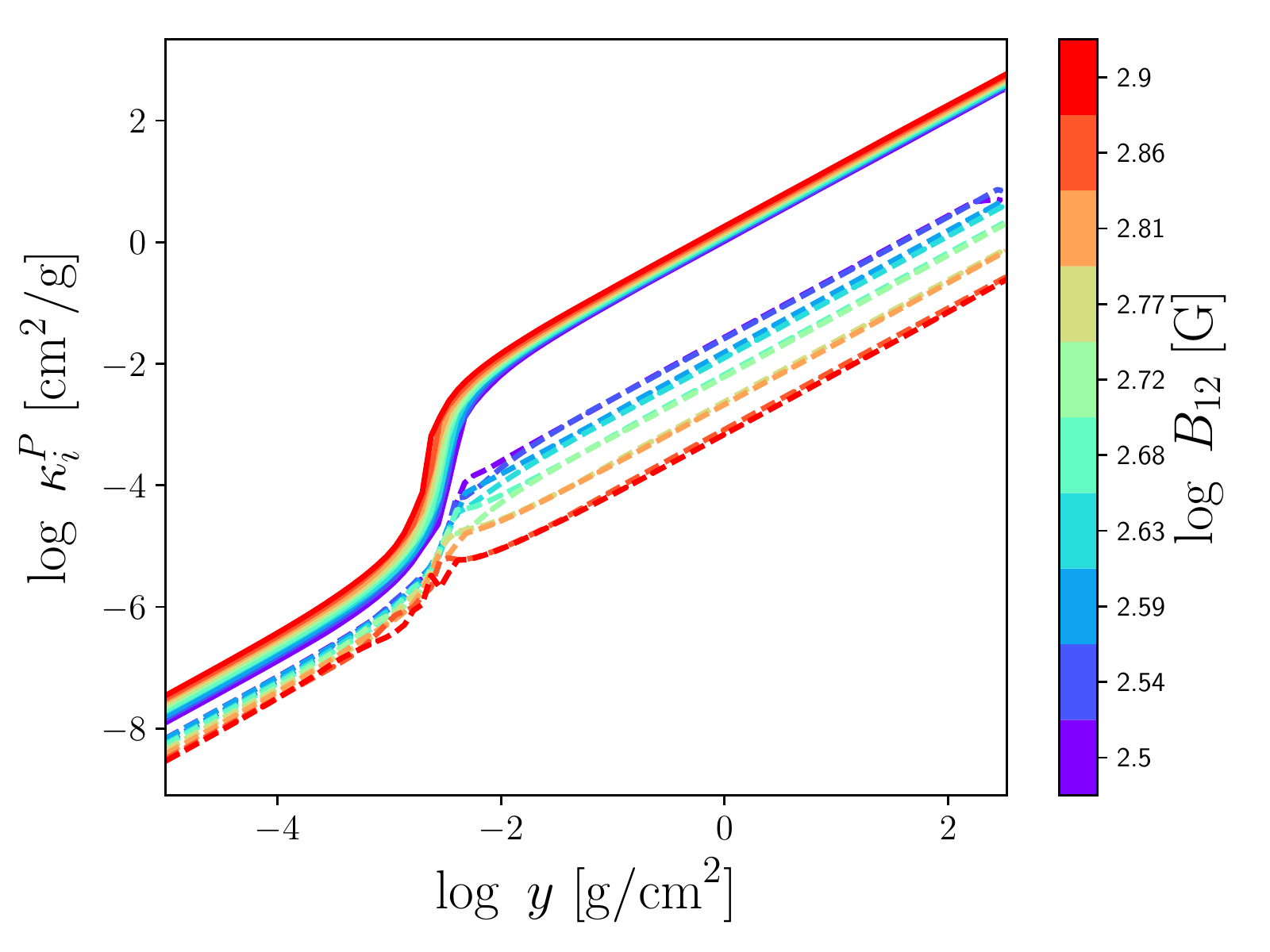}
\includegraphics[width=8.0cm,viewport=5 0 450 350,clip]{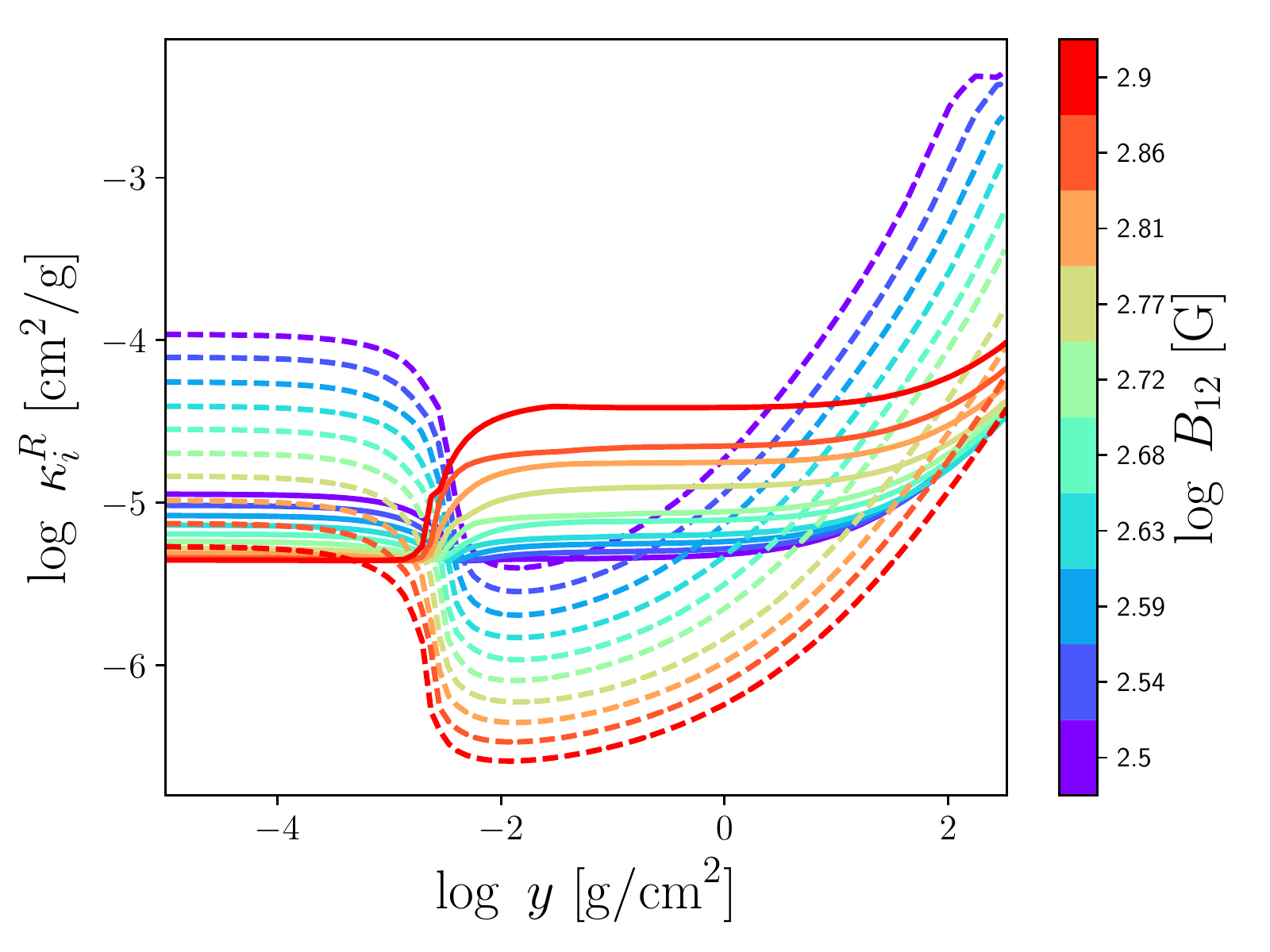}
\caption{Same as Fig. \ref{fig:seq_L}, but varying the magnetic  field  (vertical color bar)
and the stopping  column density (see Sec. \ref{fieldyo} for details).
Here, the total luminosity is set to $L_\infty=5\times10^{36}~\rm{erg/s}$.}
\label{fig:seq_B2}
\end{figure*}

\subsubsection{Models with different magnetic field}
\label{magfield}
Fig. \ref{fig:seq_B} shows  the solutions for  magnetic fields
in the range $B = 3\times 10^{14}-10^{15}$~G,  stopping
column density  $y_{_0} = 200~\rm{gr~cm}^{-2}$, and total luminosity
$L= 5\times 10^{36}~\rm{erg~cm}^{-3}$.  As shown in the top-left panel, by
varying  the magnetic field, and  thus the opacities,  no
significant changes are  produced in the temperature profile.
Therefore, this suggests that the temperature  profile is, instead,
determined just by the total luminosity  and stopping column density,
as  discussed in the previous section (Sec. \ref{colden}).

The right lower panel shows how the variation  of the magnetic
fields  introduces significant   differences in the
Rosseland mean opacities  $\kappa_1^{_R}$ and $\kappa_2^{_R}$.
However, for these opacities,
the polarization does not change monotonically with the
magnetic field. Instead, the polarization increases as the solutions approach to
both  the lower and upper end range of the magnetic field considered.

{

The behaviour of the mean opacities is now much more complicated. By increasing 
the magnetic field, the X-mode opacities decrease because of the X-mode-reduction 
factor $\sim 1/(\omega^2_{c,e}\omega)$ and $\sim (\omega/\omega_{c,e})^2$
present in the free-free and scattering opacities, respectively. However, it should
be noticed that by varying the magnetic field also changes the energy associated to the mode
switch at the vacuum resonance and its location in the atmosphere. This effect is responsible 
for {\bf a)} the fact that the curves of the various Planck mean opacities become closer to each
 other in the external atmospheric layers and {\bf b)} the different location of the crossing 
 points of the Rosseland mean opacities (for the two modes) in different models.

}

\subsubsection{Models with different magnetic field and different stopping length}
\label{fieldyo}

We  want to test a scenario in which, as expected,
the stopping column density depends on the magnetic field.
We describe  this dependency in a logarithmic scale as
$\log y_{_0} = -8.7 + 0.75 \log B$ (in cgs units), fairly compatible
with the mechanisms discussed (in Sec. \ref{stopping})
for decelerating  fast magnetospheric particles
(with Lorentz factor $\gamma = 10^3$).
{ We stress that the length scale at which the pair cascade 
propagates (produced from X-mode photons) is, for these 
values of parameters, roughly independent
 on whether the bombarding charge is stopped by a Magneto-Coulomb
 interaction or by Compton drag.}
The models are run for a total luminosity  $L_{\infty} = 5\times 10^{36}~\rm{erg~s}^{-1}$,
a value that is more convenient to ensure numerical convergence.
In addition, we also found that it is numerically convenient to solve the
the transport of radiation for luminosity in mode  2 (see eq. \ref{eq:luminosity}).
For doing that, the modes in the Rosseland and Planck
mean opacities are flip as $\kappa_i^{_R} \rightarrow  \kappa_{3-i}^{_R}$
and $\kappa_i^{P} \rightarrow  \kappa_{3-i}^{P}$  (with $i = 1, 2$), respectively,
in the numerical routines that perform their calculations.
Notice that there is no
need to perform this operation in the mode exchange scattering opacity
as $\kappa_{12} =\kappa_{21}$.

Figure \ref{fig:seq_B2}, top-left panel, shows the solution for the temperature
profile. Since lower magnetic fields produce smaller stopping column
densities, consequently,  the temperature of the  external layer increases
for lower magnetic fields (as the results discussed in  Sec. \ref{colden}).
By varying the luminosity this trend, qualitatively,
still holds (although no quantitative solutions are shown, since,
at column densities just below $\sim y_{_0}$, numerical 
oscillations start to grow in the region with the steep luminosity 
gradient).

The solutions for the luminosities in each normal mode, top-right panel, show
that the polarization decreases almost monotonically with the magnetic field
(except at the upper end of the magnetic field range).  This is at
variance with the results shown in the Sec. \ref{magfield}, 
and it shows that that the polarized signal is affected more
strongly   by the magnetic field effects 
 on the stopping  column density rather than those on the opacities.

\begin{figure}
\includegraphics[width=9.5cm,viewport=10 0 500 400,clip]{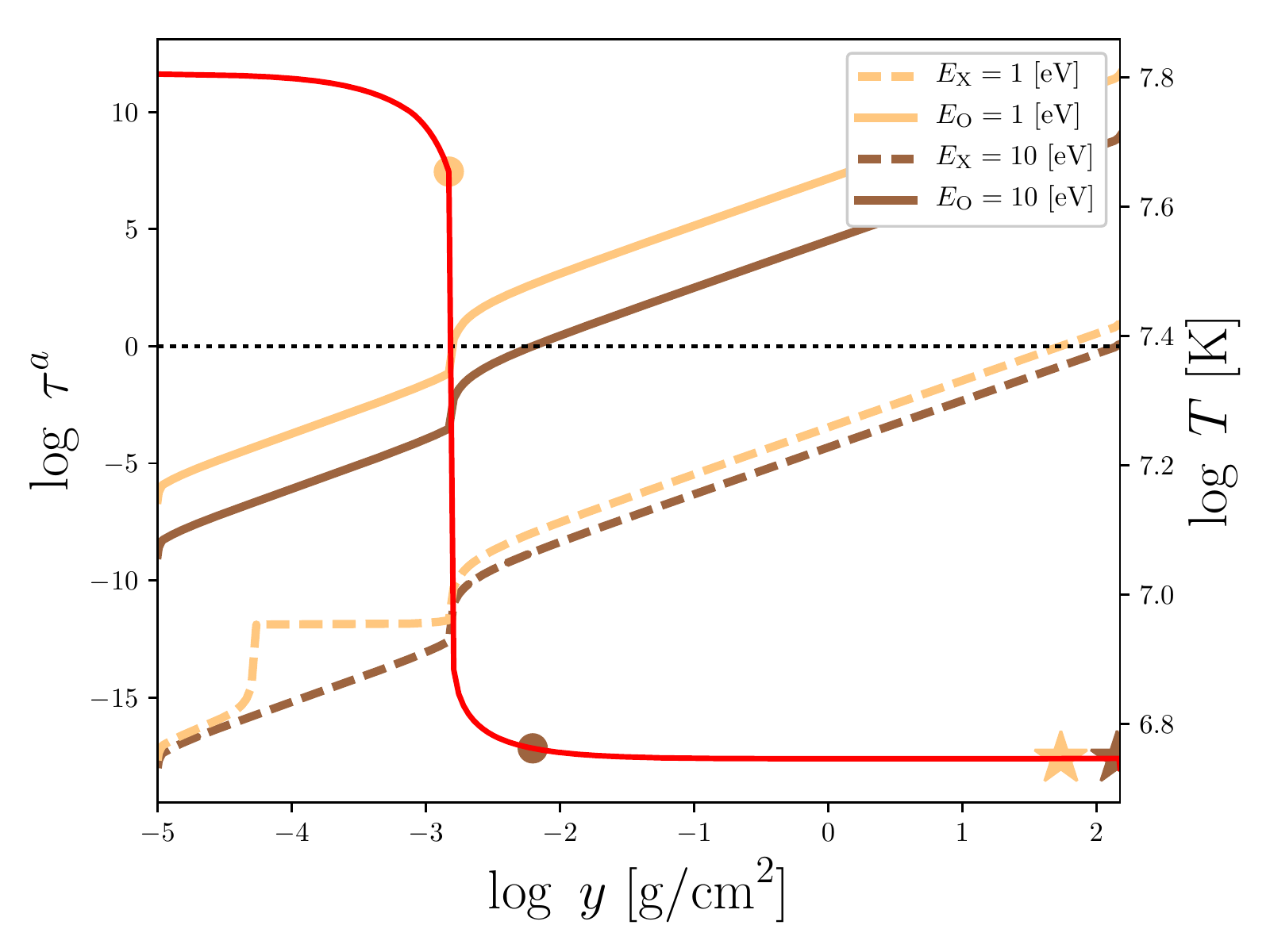}
\caption{Optical depth as a function of column density for photons in the
optical ($E=1$~eV)  and ultraviolet band ($E=10$~eV).
The model  is computed for $L_\infty = 10^{36}~\rm{erg/s}$,
$y_{_0} = 100~\rm{gr/cm}^2$ and $B=4\times 10^{14}$~G.
Solid and dashed  lines correspond to  O-mode and X-mode photons,
respectively.  The temperature profile is shown by the red solid line
(the temperature scale is on the right-hand side).
 The yellow (brown) circle and yellow (brown) star  show the
location of the  radiation-atmosphere decoupling, $\tau \sim 1$, for
O-mode and X-mode photons in the optical (ultraviolet) band,
respectively.
}
\label{fig:tau_decoupling}
\end{figure}

\subsubsection{Spectral expectations}
\label{depth}

Some information about the spectral properties of the emerging 
radiation can be obtained even in the gray model looking at the 
depth at which photons of different energies decouple.
In particular,
we compute  the energy dependent  optical depth (for optical and
soft X-rays photons) 
 as 
\begin{equation}
\tau_{_E}^{_i} = \int_{y_{min}}^{y} \kappa_{_E}^{a,i} \rm{d}y\prime\, ,
\end{equation} 
where $\kappa_{_E}^{a,i}$  (with $i=1,2$) is the energy dependent, 
angle integrated  absorption opacity (using $\tau_{_E}^{_i} =0$ at $y=y_{min}$).

Figure \ref{fig:tau_decoupling} shows the location of
the photon decoupling for the two normal modes.
In particular, for O-mode photons with energy
$E= 1$~eV, i.e. in the optical band,   we can see that they decouple
in the region of the hot layer. This suggest that the  spectrum in the optical,
and potentially  also in the infrared,
may have a flux excess due to the radiation from this hot layer.
Instead, X-mode photons with energy $E=10$~eV (in the ultraviolet
band), as well as photons with
higher energies (in both normal modes) decouple deeper in the
atmosphere. These photons are produced in regions with lower
and nearly uniform temperature, and so  their spectrum may be potentially
described  by a single black body. These results
suggest that the spectrum  from a magnetized atmosphere heated by
a particle bombardment (single black body
with optical/infrared excess), may show substantial deviations from
those produced by passively cooling neutron stars (spectrum harder than
a blackbody). Models accounting for the  full frequency
transport of radiation are definitely required to assess this issue
and will be the focus of future work.

\section{Summary and conclusions}
\label{conclusions}

We have presented numerical models for gray atmospheres
of magnetized NSs heated by  particle bombardment. We assumed a  pure, fully
ionized hydrogen composition and accounted for the polarized
transport of radiation induced by the strong magnetic field.
Our main results can be summarized as follows.

\begin{itemize}

\item[-] A hot layer is created in the most external region of
the atmosphere ($T \sim 10^8$~K), similar to what it has been
found in previous studies of atmospheres heated by accretion 
\citep{Turolla94,Zane00}.  We found that, in the case presented here,
the spatial extent of the layer is primary  determined  by the
luminosity and the stopping column density  for the particle bombardment.
If there is no significant contribution  from
the NS cooling (i.e. the luminosity vanishes in the internal
layer, as assumed here)  then the internal regions of the atmosphere are
characterized  by a lower (below $T\sim 10^{7}$~K), uniform temperature
that is primary determined by the total model luminosity.

\item[-] As a consequence, we expect that in the range of frequencies
that decouple in the deeper atmospheric layers (e.g. above $\sim 10$~eV)
the emergent spectrum may be  well approximated  by a single
blackbody. Instead,  photons   that decouple in the hot layer
(below $\sim 1$~eV)  may produce an optical/infrared
flux excess, significantly above the extrapolation of the blackbody
spectrum coming from the atmosphere interior (see \citealt{Zane00}
for similar discussions).

\item[-] The expected degree of linear polarization in the emergent
spectrum is correlated with the luminosity and anticorrelated
with the stopping column density (and almost similarly with the magnetic
field, see text for details). Depending on these parameters, the expected
intrinsic linear polarization, defined here as  $PF = (L_2 -L_1)/(L_2 + L_1)$,
for  a  single patch  in the atmosphere, may vary between  $0.4 < PF < 0.9$.

\item[-]  Therefore, variations in the intrinsic  polarization
may be present in the thermal emission from transient magnetar, as the
 luminosity in these sources can decrease by orders of magnitude.
\end{itemize}

Our results have been obtained under a number of assumptions
and simplifications.
A major simplification is, the assumption of pure H and fully
ionized  atmosphere. Strong magnetic fields, $B=10^{14}-10^{15}$~G,
increase  significantly  the binding energy of the atoms, and
this can lead to  additional contributions  to  the opacities.
From Fig. 3 of  \cite{Potekhin04b},  we can see that,  for the
parameters typical of the  models studied in our work,
partial ionization can be safely neglected in   the most
external (hot) layers of the  atmosphere,  where
the atomic fraction should be much lower than $0.1\%$. Although
for intermediate atmospheric  layers  the expected atomic
fraction should be $\sim 0.1 -  1\%$,  this may be still  important
 as  bound-bound and bound-free transitions (particularly in the
 energies range $E=1-10$~keV) may change the opacities by orders of
 magnitude (see e.g. Fig. 4 and 5 in \citealt{Ho03b}).
In addition, strong magnetic fields can also induce  the formation of
complex molecules. However, \cite{Potekhin04a}  showed that the formation of
those  molecules can be suppressed  for   effective temperatures  slightly
above $T_{\rm{eff}} \sim 4 \times 10^5 (B/10^{14}~\mathrm{G})^{1/4}$.  Therefore this
effect is not expected to be relevant  in the range  of $T$ and $B$ studied
in this work.

Furthermore, vacuum polarization and mode conversion
induced by the strong magnetic  field can also  produce substantial effect
on the opacities. For example, \cite{Ho03a} studied two  simplified limiting
cases: full mode conversion and no-mode conversion, showing that both cases
can lead to large polarization in the spectrum, i.e., almost $\sim 100\%$
in the soft X-ray  band.  However, as also pointed out by \cite{Ho03a} the
effect is not well understood yet and  a realistic scenario may lie
between the two limiting cases. In spite of
the crude expression used here to estimate the polarization degree,
our work shows that intermediate  cases for mode  conversion can produce
a much smaller (and luminosity-dependent) degree of  polarization. However,
it should be also noticed  that non-orthogonality of the normal modes
may affect a substantial range of photon propagation directions at the
vacuum MCP.  In these cases, a proper calculation
of the transport of radiation should be performed via the evolution of
the Stokes parameters. The mode conversion effect basically tends to
decrease the difference  between the opacity of the two normal modes, and
therefore the way in which this effect is treated affects critically
the expected degree of polarization. Further work on this important
issue needs therefore to be carried on in order to produce more quantitative
predictions. Future  observations of magnetars  with X-ray polarimeters (IXPE,
\citealt{Weisskopf13}; eXTP, \citealt{Zhang16}) may provide crucial
information in directing the theoretical effort.

We point out that the calculation of the stopping column density
is also affected by additional uncertainties. In particular,    the cross section for
magneto-Coulomb interaction, photon splitting (either induced by the magnetic
field or the coulomb field of a nucleus) and  scattering cross sections
used in literature and in this work are valid for $B < 10^{14}$~G, i.e.,
below the range of field strengths we considered. This affects
directly the calculation of the energy deposition in the atmospheric layer. { To
mitigate this, we considered a large range of stopping 
column density values, $65-500~\mathrm{g~cm}^{-2}$.  
We also caveat that across the paper we have considered a representative 
value 
$\gamma=10^{3}$ for the Lorentz factor of the (monoenergetic) 
bombarding 
electrons, but the problem of the determination of the spectrum of the 
impinging charges is far from being understood.  
In principle this is a reasonable first guess if the 
Lorentz factor of the charges near the star surface is kept around the 
threshold for efficient resonant Compton drag to occur, since it 
is $\gamma \sim B/\Theta \sim 10^3$ for $B \sim 10^{14}$~G and $T \sim 
1$~keV. However, the value of
this Lorentz factor depends on the magnetic field and photon temperature  
at the surface, so it may vary considerably. If the Lorentz factor of the 
charges is as high as $\sim 10^6-10^7$, electrons are still efficiently 
stopped by magneto-Coulomb collisions (which length scale depends weakly 
on $\gamma$, see Eq.~2), but the resulting photon can be much more 
energetic and in turn the resulting pair cascade can penetrate much 
further, up to $\sim 700-1000~\mathrm{g~cm}^{-2}$. 

In principle, 
the atmosphere may also modulate its penetration depth and 
atmospheric profile to ultrarelativistic electrons. The exploration of 
possible equilibrium solutions that account for the backreaction of the 
atmosphere would require to modify our scheme 
and calculate (numerically) $y_0$ either assuming an approximated scaling with the temperature, as e.g. in 
\cite{Baring11}, or solving the full kinetic particle transport.
This is, however, outside the scope of this first investigation.
Our results have shown that, when treating $y_0$ as a model parameter, the 
main effects
of varying the stopping column length are actually restricted in the most
external layers. Therefore, our conclusion regarding the 
creation of a blackbody like spectrum from the most internal
atmospheric layers should hold, as it showed to be pretty
much independent of the stopping column density.
}

\section*{Acknowledgements}

DGC acknowledges the financial support  by CONICYT
Becas-Chile fellowship (No. 72150555). 
We are grateful to an anonymous referee for a careful
 reading and constructive comments on the paper


\bibliographystyle{mnras}

\bibliography{biblio}







\bsp    
\label{lastpage}
\end{document}